\documentclass[10pt]{elsart}
\usepackage{amsmath}
\usepackage{graphicx}
\usepackage{axodraw}
\usepackage{feynmp}
\def\Order{\mathcal{O}}
\def\noeqnumlb{\notag\\[1.5ex]}

\def\yoff{-3}
\def\dash{2}    % Dash size for massless lines
\def\vtxsz{1.2} % Size of vertex
\SetScale{1}    % Scale factor of graphic
\SetWidth{0.7}  % Line width

\def\TadA{\begin{picture}(40,40)(0,\yoff)\CArc(3.83,6)(6,101.25,318.75)\CArc(36.17,6)(6,-138.75,78.75)\CArc(20,34)(6,-18.75,198.75)\Line(20,15.33)(14.32,32.07)\Line(20,15.33)(25.68,32.07)\Line(20,15.33)(8.35,2.04)\Line(20,15.33)(2.66,11.88)\Line(20,15.33)(37.34,11.88)\Line(20,15.33)(31.65,2.04)\Vertex(20,15.33){\vtxsz}\end{picture}}
\def\TadB{\begin{picture}(40,40)(0,\yoff)\CArc(20,20)(20,90,270)\CArc(20,20)(20,-90,90)\DashCArc(46.67,20)(33.33,143.13,216.87){\dash}\DashCArc(-6.67,20)(33.33,-36.87,36.87){\dash}\Vertex(20,0){\vtxsz}\Vertex(20,40){\vtxsz}\end{picture}}
\def\TadC{\begin{picture}(40,40)(0,\yoff)\CArc(20,20)(20,90,270)\CArc(20,20)(20,-90,90)\CArc(46.67,20)(33.33,143.13,216.87)\CArc(-6.67,20)(33.33,-36.87,36.87)\Vertex(20,0){\vtxsz}\Vertex(20,40){\vtxsz}\end{picture}}
\def\SEA{\begin{picture}(60,40)(-5,\yoff)\Line(-5,20)(10,20)\Line(40,20)(55,20)\DashCArc(25,13.75)(16.25,22.62,157.28){\dash}\DashCArc(25,26.25)(16.25,-157.28,-22.62){\dash}\Vertex(10,20){\vtxsz}\Vertex(40,20){\vtxsz}\CArc(5,28)(5,0,240)\Line(10,20)(10,28)\Line(10,20)(2.5,23.67)\CArc(45,28)(5,-60,180)\Line(40,20)(40,28)\Line(40,20)(47.5,23.67)\end{picture}}
\def\SEB{\begin{picture}(90,40)(-5,\yoff)\Line(-5,20)(10,20)\Line(70,20)(85,20)\DashCArc(25,13.75)(16.25,22.62,157.28){\dash}\DashCArc(25,26.25)(16.25,-157.28,-22.62){\dash}\DashCArc(55,13.75)(16.25,22.62,157.28){\dash}\DashCArc(55,26.25)(16.25,-157.28,-22.62){\dash}\CArc(75,28)(5,-60,180)\Line(70,20)(70,28)\Line(70,20)(77.5,23.67)\Vertex(10,20){\vtxsz}\Vertex(40,20){\vtxsz}\Vertex(70,20){\vtxsz}\end{picture}}
\def\SEC{\begin{picture}(60,40)(-5,\yoff)\Line(-5,20)(10,20)\Line(40,20)(55,20)\DashCArc(25,13.75)(16.25,22.62,157.28){\dash}\DashCArc(25,26.25)(16.25,-157.28,-22.62){\dash}\DashLine(10,20)(40,20){\dash}\CArc(45,28)(5,-60,180)\Line(40,20)(40,28)\Line(40,20)(47.5,23.67)\Vertex(10,20){\vtxsz}\Vertex(40,20){\vtxsz}\end{picture}}

\begin{document}

\begin{frontmatter}
\begin{flushleft}
TTP07-32\\
SFB/CPP-07-76\\
arXiv:0711.2636
\end{flushleft}
\title{Higher Moments of Heavy Quark Correlators in the Low Energy Limit at
  $\mathcal{O}(\alpha_s^2)$} 
\author[Karlsruhe]{A.~Maier}, 
\author[Karlsruhe]{P.~Maierh\"ofer} and 
\author[Karlsruhe]{P.~Marquard}
\address[Karlsruhe]{Institut f\"ur Theoretische Teilchenphysik,
Universit\"at Karlsruhe, 76128 Karlsruhe, Germany}
\begin{abstract}
We present the first 30 moments of the low energy expansions of the
vector, axial-vector, scalar and pseudo-scalar heavy quark correlation
functions at three-loop order, including the singlet contribution which
appears for the first time at three loops. In addition we compare the
behavior of the moments for large $n$ with the prediction from threshold
calculations.  
\end{abstract}
\end{frontmatter}

% ================== %
\section{Introduction}
% ================== %

Correlators of two currents $j_{[\mu]}(x)$, where $j_{[\mu]}(x)$ is the
scalar $\bar{\psi}\psi$, pseudo-scalar $i\bar{\psi}\gamma_5\psi$, vector
$\bar{\psi}\gamma_\mu\psi$, or axial-vector
$\bar{\psi}\gamma_\mu\gamma_5\psi$ current, are important for many
interesting phenomenological purposes.  Their low energy expansions, in
particular, are of interest for applications based on sum rules
\cite{Shifman:1978by,Reinders:1984sr} like the determination of the
charm and bottom quark mass from $R(s)$ \cite{Kuhn:2007vp,Kuhn:2001dm}.

In three-loop approximation, i.\,e.\ $\mathcal{O} (\alpha_s^2)$, the
moments of the Taylor expansion for the four correlators were evaluated
up to $(q^2)^8$ in \cite{Chetyrkin:1997mb} using recursion algorithms
originally suggested in \cite{Broadhurst:1991fi}. The series
was evaluated in four-loop approximation up to
the first physical moment $(q^2)^1$ 
\cite{Chetyrkin:2006xg,Boughezal:2006px} employing the Laporta algorithm
\cite{Laporta:2001dd}. The three-loop singlet
contributions have been calculated in \cite{Chetyrkin:1998ix} up to the
seventh and eighth moments for the axial-vector and the scalar and
pseudo-scalar correlators, respectively.

Recently, using a completely different technique based on master
differential equations \cite{Caffo:1998du,Remiddi:1997ny}, the three-loop
vector correlator has been calculated up to $(q^2)^{30}$
\cite{Boughezal:2006uu}. 

Higher moments of the  pseudo-scalar correlator might be used to
determine the impact of non-perturbative effects in the determination of
quark masses from lattice calculations \cite{Lepage}.

In this work we calculate the low energy expansions for the four types
of correlators in order $\alpha_s^2$ up to $(q^2)^{30}$, including the
logarithms which arise from the singlet contribution, given by double
triangle diagrams. 

This paper is organized as follows: In Section \ref{sec:met} we briefly
present the methods used in the calculation, paying special attention to
the singlet parts. The results are presented in Section \ref{sec:sc}
where we also compare the explicitly calculated moments with the
behavior for large powers of $q^2$ derived from the threshold region.
Section \ref{sec:conclusion} contains a brief summary and conclusions.

% ============================ %
\section{Definitions and Methods}\label{sec:met}
% ============================ %

The polarization functions are defined by
 \begin{eqnarray}
   (-q^2 g_{\mu\nu} + q_\mu q_\nu) \Pi^\delta (q^2) + q_\mu q_\nu
   \Pi_L^\delta(q^2) &=& i\int dx e^{iqx} \langle 0 |Tj_\mu^\delta(x)
   j_\nu^\delta(0)|0\rangle \\
   &&\hspace{12ex}\mbox{for} \quad \delta=v,a, \nonumber \\
   q^2 \Pi^\delta(q^2) &=& i \int dx e^{iqx} \langle 0 |Tj^\delta(x)
   j^\delta(0)|0\rangle \\
   &&\hspace{12ex}\mbox{for} \quad \delta=s,p, \nonumber
 \end{eqnarray}
with the currents 
\begin{equation}
  j_\mu^v = \bar\psi \gamma_\mu \psi, \quad j_\mu^a = \bar\psi
  \gamma_\mu \gamma_5 \psi, \quad j^s = \bar \psi\psi, \quad j^p =
  i \bar\psi\gamma_5 \psi. 
\end{equation}
In the low energy limit the polarization functions can be written as a
series in $z=\frac{q^2}{4 m^2}$, where $m$ is the mass of the heavy quark,
\begin{equation}
  \Pi^\delta(q^2) = \frac{3}{16\pi^2} \sum_{n>0} C^\delta_n z^n ,
\end{equation}
with
\begin{eqnarray}
  C^\delta_n &=& C^{(0),\delta}_n + \frac{\alpha_s}{\pi} C_F
  C^{(1),\delta}_n + \left (\frac{\alpha_s}{\pi} \right )^2
  C^{(2),\delta}_n + \cdots,\\
  C^{(2),\delta}_n &=& C_F^2 C_{A,n}^{(2),\delta} + C_F C_A
  C_{NA,n}^{(2),\delta} + C_F T_F n_l C_{l,n}^{(2),\delta} \nonumber
   \\&& \qquad
+ C_F T_F n_h C_{F,n}^{(2),\delta} + C_F T_F n_h
   C_{S,n}^{(2),\delta} .
\end{eqnarray}

\begin{figure}
\begin{fmffile}{fullsingdias}
\fmfset{arrow_len}{2mm}
\fmfset{dot_size}{thick}
\fmfset{curly_len}{1.5mm}
\fmfpen{thin}
\begin{fmfshrink}{.75}
\begin{equation*}
\begin{split}
%j_{\mu\nu,S} = &
\quad
\parbox{75pt}{
  \begin{fmfgraph*}(75,75)
   \fmfleft{i1}
   \fmfright{o1}
   \fmftop{t1,t2}
   \fmfbottom{b1,b2}
   \fmf{photon}{i1,v1}
   \fmf{photon}{v2,o1}
   \fmf{fermion,tension=.3}{v1,v3,v4,v1}
   \fmf{fermion,tension=.3}{v2,v5,v6,v2}
   \fmf{gluon,tension=1}{v3,v5}
   \fmf{gluon,tension=1}{v4,v6}
   \fmf{phantom}{t1,v3}
   \fmf{phantom}{t2,v5}
   \fmf{phantom}{b1,v4}
   \fmf{phantom}{b2,v6}
   \fmfdotn{v}{6}
  \end{fmfgraph*}
}
\quad
%-\quad 2
\quad
\parbox{75pt}{
  \begin{fmfgraph*}(75,75)   
    \fmfleft{i1}
   \fmfright{o1}
   \fmftop{t1,t2}
   \fmfbottom{b1,b2}
   \fmf{photon}{i1,v1}
   \fmf{photon}{v2,o1}
   \fmf{fermion,tension=.3}{v1,v3,v4,v1}
   \fmf{dashes_arrow,tension=.3}{v2,v5,v6,v2}
   \fmf{gluon,tension=1}{v3,v5}
   \fmf{gluon,tension=1}{v4,v6}
   \fmf{phantom}{t1,v3}
   \fmf{phantom}{t2,v5}
   \fmf{phantom}{b1,v4}
   \fmf{phantom}{b2,v6}
   \fmfdotn{v}{6}
  \end{fmfgraph*}
}
\quad
%+
\quad
\parbox{75pt}{
  \begin{fmfgraph*}(75,75)   
    \fmfleft{i1}
   \fmfright{o1}
   \fmftop{t1,t2}
   \fmfbottom{b1,b2}
   \fmf{photon}{i1,v1}
   \fmf{photon}{v2,o1}
   \fmf{dashes_arrow,tension=.3}{v1,v3,v4,v1}
   \fmf{dashes_arrow,tension=.3}{v2,v5,v6,v2}
   \fmf{gluon,tension=1}{v3,v5}
   \fmf{gluon,tension=1}{v4,v6}
   \fmf{phantom}{t1,v3}
   \fmf{phantom}{t2,v5}
   \fmf{phantom}{b1,v4}
   \fmf{phantom}{b2,v6}
   \fmfdotn{v}{6}
  \end{fmfgraph*}
}
%  \\
% +&\quad
% \parbox{75pt}{
%   \begin{fmfgraph*}(75,75)
%    \fmfleft{i1}
%    \fmfright{o1}
%    \fmftop{t1,t2}
%    \fmfbottom{b1,b2}
%    \fmf{photon}{i1,v1}
%    \fmf{photon}{v2,o1}
%    \fmf{fermion,tension=.3}{v1,v3,v4,v1}
%    \fmf{fermion,tension=.3}{v2,v6,v5,v2}
%    \fmf{gluon,tension=1}{v3,v5}
%    \fmf{gluon,tension=1}{v4,v6}
%    \fmf{phantom}{t1,v3}
%    \fmf{phantom}{t2,v5}
%    \fmf{phantom}{b1,v4}
%    \fmf{phantom}{b2,v6}
%    \fmfdotn{v}{6}
%   \end{fmfgraph*}
% }
% \quad
% -\quad 2
% \quad
% \parbox{75pt}{
%   \begin{fmfgraph*}(75,75)   
%     \fmfleft{i1}
%    \fmfright{o1}
%    \fmftop{t1,t2}
%    \fmfbottom{b1,b2}
%    \fmf{photon}{i1,v1}
%    \fmf{photon}{v2,o1}
%    \fmf{fermion,tension=.3}{v1,v3,v4,v1}
%    \fmf{dashes_arrow,tension=.3}{v2,v6,v5,v2}
%    \fmf{gluon,tension=1}{v3,v5}
%    \fmf{gluon,tension=1}{v4,v6}
%    \fmf{phantom}{t1,v3}
%    \fmf{phantom}{t2,v5}
%    \fmf{phantom}{b1,v4}
%    \fmf{phantom}{b2,v6}
%    \fmfdotn{v}{6}
%   \end{fmfgraph*}
% }
% \quad
% +
% \quad
% \parbox{75pt}{
%   \begin{fmfgraph*}(75,75)   
%     \fmfleft{i1}
%    \fmfright{o1}
%    \fmftop{t1,t2}
%    \fmfbottom{b1,b2}
%    \fmf{photon}{i1,v1}
%    \fmf{photon}{v2,o1}
%    \fmf{dashes_arrow,tension=.3}{v1,v3,v4,v1}
%    \fmf{dashes_arrow,tension=.3}{v2,v6,v5,v2}
%    \fmf{gluon,tension=1}{v3,v5}
%    \fmf{gluon,tension=1}{v4,v6}
%    \fmf{phantom}{t1,v3}
%    \fmf{phantom}{t2,v5}
%    \fmf{phantom}{b1,v4}
%    \fmf{phantom}{b2,v6}
%    \fmfdotn{v}{6}
%   \end{fmfgraph*}
% }
\end{split}
\end{equation*}
\end{fmfshrink}
\end{fmffile}
\label{fig:dias}
\caption{Diagrams for the singlet contribution. The first diagram
  contributes to scalar, pseudo-scalar and axial-vector correlators,
  while the latter two are only taken into account for the
  axial-vector to cancel the anomaly.  
 Solid and dashed lines 
  denote massive and massless lines, respectively.}
\end{figure}
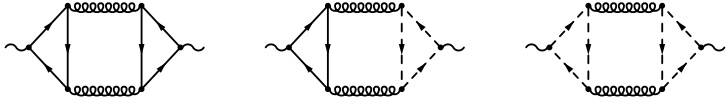

The singlet contributions $C_{S,n}^{(2),\delta}$ arise from the diagrams
shown in Fig. \ref{fig:dias}. It has to be noted that these are not 
simple power series in $z$ but contain logarithms of the form
$\log (-\frac{q^2}{m^2})$. This logarithmic dependence on $q^2$ reflects the
presence of massless cuts in the diagrams.

The expansion in the 
limit $q^2 \to 0$  can be achieved in two different ways. One
possibility is to first expand the propagators and later reduce
the resulting massive tadpoles to master integrals. For each order in
the $z$-expansion two more powers of the propagators arise.
For large $n$ this makes the reduction to master integrals through
the reduction formalisms of Broadhurst or Laporta very cumbersome. 
 The second possibility 
is the  reduction of the full propagators to master integrals. 
These master
integrals have then to be known at least as an expansion in the
external momentum which can be calculated efficiently to
high orders. The price to pay  is the increased
difficulty in obtaining the necessary reduction to master integrals.   
In
this paper we choose the latter approach, which we present in more
detail in the following.

In the first step, the diagrams are generated with
\texttt{QGRAF} \cite{Nogueira:1991ex}. The occurring topologies are
identified with the help of \texttt{q2e} and \texttt{exp}
\cite{Seidensticker:1999bb,Harlander:1997zb}. 
The integrals are reduced to scalar 
propagator-type integrals of the form
\begin{equation}
 P(q^2,a_1,\dots,a_9) = \int\frac{[dk_1][dk_2][dk_3]}{D_1^{a_1}D_2^{a_2}\dots D_9^{a_9}}\quad. \label{propagatorintegral}
\end{equation}
using the computer algebra program \texttt{FORM} \cite{Vermaseren:2000nd}.
The denominators are given by $D_i=(l-m_i^2+i\varepsilon)$, where $l$ is
a linear combination of the loop momenta $k_i$ and the external momentum
$q$, and $m_i\in\{0,m\}$ is the mass of the corresponding propagator.
Only one non-zero mass is taken into account for the heaviest quark,
lighter quarks are treated as massless. The powers $a_i$ of the
denominators are integers. Irreducible scalar products are expressed as
propagators with negative powers. 
The integrals are dimensionally regularized
in space-time dimension $d=4-2\,\epsilon$.

In a next step, Integration-by-Parts \cite{Chetyrkin:1981qh} and
Laporta's algorithm \cite{Laporta:2001dd,Laporta:1996mq} are used to
reduce all needed integrals of the form (\ref{propagatorintegral}) to
master integrals. This reduction is done with the program
\texttt{Crusher} \cite{crusher}. \texttt{Crusher} uses
\texttt{GiNaC} \cite{bauer-2002-33} for algebraic manipulations and
\texttt{Fermat} \cite{fug} for the simplification of the intermediate
expressions using a special interface \cite{tentyukov-2007-176}.  As a
result of the reduction one finds a total of 55 master
integrals. 

\begin{figure}[h]
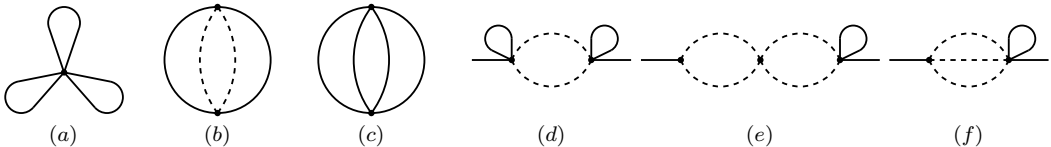

  \vspace{5mm}
  \begin{center}
    $\underset{(a)}{\TadA}$\hspace{5mm}
    $\underset{(b)}{\TadB}$\hspace{5mm}
    $\underset{(c)}{\TadC}$\hspace{5mm}
    $\underset{(d)}{\SEA}$
    $\underset{(e)}{\SEB}$
    $\underset{(f)}{\SEC}$
    \caption{Diagrams, which are chosen as boundary conditions for the
      expansion of the propagator-type master integrals in
      $q^2$. The diagrams $(d)-(f) $ contribute only to the singlet
      part.\label{boundaries}}
  \end{center}
\end{figure}

The master integrals have to be expanded for small external momenta
around $q^2=0$. To achieve this, we use the scaling equation
\begin{equation}
  \left(q^2\frac{\partial}{\partial q^2} + m^2\frac{\partial}{\partial m^2} - \hat{D}\right)M_i(q^2) = 0
\end{equation}
for the master integrals $M_i$. $\hat{D}$ applied to $M_i$ gives the
mass dimension of $M_i$. Carrying out the mass derivative produces
integrals with additional powers of propagators, that are again mapped
to master integrals. This gives a system of coupled inhomogeneous linear
differential equations in $q^2$. For the non-singlet part, the system is
solved by a Taylor series
\begin{equation}
M_i(q^2) = \sum_{k=0}^\infty M_i^{(k)}(q^2)^k\quad,
\end{equation}
where the coefficients $M_i^{(k)}$ have to be calculated.
Diagrammatically, the coefficients are tadpole diagrams. As boundary
conditions, the three-loop master tadpoles depicted in
Fig. \ref{boundaries} $(a)$-$(c)$ are chosen. The $M_i^{(k)}$ can be
expressed as linear combinations of these. For the singlet part,
non-integer powers of $q^2$ have to be taken into account:
\begin{equation}
  M_i(q^2) = \sum_{k} \left(M_i^{(k)}+M_{i,s}^{(k)}(q^2)^{-\epsilon}+M_{i,ss}^{(k)}(q^2)^{-2\epsilon}\right)(q^2)^k\quad.
\end{equation}
In this case, the boundary conditions for the coefficients
$M_{i,s}^{(k)}$ and $M_{i,ss}^{(k)}$ are the diagrams depicted in
Fig. \ref{boundaries} $(d)$ and $(e),(f)$, respectively. Generation
and solution of the system are implemented in a \texttt{Mathematica}
program. 
With our setup we can calculate the $q^2$ expansion to almost
arbitrary depths. In this work we limit ourselves to the first $30$
moments. This way of calculating expansions of integrals was proposed
in \cite{Caffo:1998du,Remiddi:1997ny} and applied to the vector
current in \cite{Boughezal:2006uu}.

To perform the renormalization we need the strong coupling  
$\alpha_s$ \cite{Tarasov:1980au} and the mass renormalization constant at two-loop
order \cite{Gray:1990yh,Tarasov:1982gk,Gorishnii:1983zi}. 

For the pseudo-scalar and axial-vector currents we have to pay special
attention to the treatment of $\gamma_5$. While in the non-singlet
contributions $\gamma_5$ can simply be considered as anticommutating
this is not possible for the singlet contributions. In the latter case
the $\gamma_5$ matrix cannot easily be removed from the problem and has
to be treated properly to avoid problems with dimensional
regularization. Therefore we follow the prescription proposed by Larin
\cite{Larin:1993tq} and use the definition $\gamma_5=\frac{i}{4!}
\epsilon_{\mu\nu\rho\sigma}\gamma^\mu\gamma^\nu\gamma^\rho\gamma^\sigma$.
Since the $\epsilon$-tensor is intrinsically a four-dimensional object
it can not be used in dimensional regularization and has to be taken out
of the calculation until the renormalization has been performed. Taking
the $\epsilon$-tensors out of the calculation the tensor structure of
the pseudo-scalar and axial-vector correlators can be cast into the form
\begin{eqnarray} {\Pi^{p}}^{\mu\nu\rho\sigma}_{\mu'\nu'\rho'\sigma'}
  &=& \Pi^{p}_1
  g^{[\mu}_{[\mu'}g^{\nu}_{\nu'}g^{\rho}_{\rho'}g^{\sigma]}_{\sigma']} +
  \Pi^{p}_2
  q^{[\mu}q_{[\mu'}g^{\nu}_{\nu'}g^{\rho}_{\rho'}g^{\sigma]}_{\sigma']}
  \ ,
\\
  {\Pi^{a}}^{\mu\nu\rho}_{\mu'\nu'\rho'} &=&\Pi^{a}_1 
  g^{[\mu}_{[\mu'}g^{\nu}_{\nu'}g^{\rho]}_{\rho']} +
  \Pi^{a}_2 q^{[\mu}q_{[\mu'}g^{\nu}_{\nu'}g^{\rho]}_{\rho']}  \ ,
\end{eqnarray}
where $[\ldots]$ denotes total antisymmetrization. 
After renormalization these expressions can be multiplied with
the $\epsilon$-tensors to obtain the final result. The correlators are
then given by
\begin{eqnarray}
  \Pi^{p} &=& \Pi^{p}_1 + \Pi^{p}_2 \ ,\\
  \Pi^{a} &=& \Pi^{a}_1  + \Pi^{a}_2\ ,\\
  \Pi_L^{a} &=& \Pi^{a}_2  \ .
\end{eqnarray}
The pseudo-scalar and the longitudinal part of the axial-vector
correlator are connected through a Ward identity.
\begin{equation}
q^2 \Pi_L^{a} (q^2) = 4 m^2 (\Pi^{p} (q^2) - q^2 (\partial
\Pi^{p} (q^2) / \partial q^2  )|_{q^2=0})
.
\label{eq:1}
\end{equation}
In order to retain this Ward identity for the singlet contribution it
is necessary to cancel the axial-vector anomaly.  For this reason we
computed the singlet part $\Pi_{\mu\nu,S}$ of the axial-vector
correlator for an isospin doublet $(\psi,\chi)$, where $\psi$ is taken
to be heavy and $\chi$ light. The diagrams contributing to the full
singlet part are depicted in Fig. \ref{fig:dias}. Note that the completely
massless diagram does only contribute to the leading moment.

% =========================== %
\section{Results and asymptotic behavior for large $n$}\label{sec:sc}
% =========================== %

The numerical results\footnote{While we agree in general with the
  results given in \cite{Boughezal:2006uu}, we disagree with several
  single digits.} for the first 30 moments of the various currents are
listed in Appendix \ref{sec:app:num}. The results are given in both the
$\overline{\mathrm{MS}}$ and the onshell scheme since both schemes have
their own range of applications.  For convenience we set the
renormalization scale $\mu = m$. Since the longitudinal part of the
axial-vector correlator can be easily obtained from the pseudo-scalar
one by use of the Ward identity (\ref{eq:1}) we give only the results
for the transversal part in this case.  

The results for the singlet contributions to the scalar, pseudo-scalar
and axial-vector correlators can be  split into a  constant and
a logarithmic part proportional to $\log(-\frac{q^2}{m^2})$. These are
denoted by $ C_{S,n}^{(2),\delta} [1]$ and $C_{S,n}^{(2),\delta} [L]$,
respectively. In the case of  the axial-vector correlator the singlet
part includes contributions from heavy-light diagrams as explained in
the previous section.

  The corresponding analytical
expressions for moments 9-12 in the $\overline{\mathrm{MS}}$ scheme can
be found in the Appendix \ref{sec:app:ana}.  The analytical results for all calculated
moments both in the $\overline{\mathrm{MS}}$ and onshell scheme
including all logarithms are available in computer readable
form from \\
\texttt{http://www-ttp.particle.uni-karlsruhe.de/Progdata/ttp07/ttp07-32}.

For large expansion depths, the moments can be compared to the
asymptotic behavior, which follows from the threshold behavior of heavy
quark production from $e^ +e^-$.
\begin{equation}
  C_{k,n} = \frac{4}{9}\int_{4m^2}^\infty\frac{ds}{s}\,R_{k}(s)\left(\frac{4m^2}{s}\right)^n = \frac{4}{9}\int_0^1 d(\beta^2)\,R_{k}(\beta)\big(1-\beta^2\big)^{n-1} \ , \label{vton}
\end{equation}
where
\begin{equation}
  k=A,NA,l,F\quad\mbox{and}\quad{}\beta=\sqrt{1-\frac{4m^2}{s}} \ .
\end{equation}

In view of the smooth behavior of the calculated moments, we expect good
agreement for $n \sim 30$. Large $n$ expansions can be found in ref.
\cite{Chetyrkin:1996ci} and derived from
\cite{Czarnecki:1997vz,Chetyrkin:1997mb,Chetyrkin:1996cf} by using eq.
(\ref{vton}). A power $\beta^j$ of the velocity in the $R$-ratio
$R_{k}(\beta)$ corresponds to a power $n^{-1-j/2}$ of the large $n$
expansion. 
%We collect here the known terms in the large $n$ expansion
%for the vector correlator at three loops.
\begin{eqnarray}
  C_{A,n}^{(2),v} &=& \frac{\pi^{9/2}}{6}n^{-1/2} - 4\,\pi^2\,n^{-1} + \frac{\sqrt{\pi}}{144}\big(23\pi^4 + 8\pi^2(6H_{n+\frac{1}{2}} \nonumber \\
  && {} - 47+36\log(2))+36(39-4\zeta_3)\big)n^{-3/2} + \Order(n^{-2}) , \nonumber \\
  C_{NA,n}^{(2),v} &=& \frac{\pi^2}{36}\big(33H_n+31-66\log(2)\big)n^{-1} + \frac{\sqrt{\pi}}{72}\big(\pi^2(36H_{n+\frac{1}{2}} \nonumber \\
  && {} + 107-120\log(2))-302-468\zeta_3\big)n^{-3/2} + \Order(n^{-2}) , \nonumber \\
  C_{l,n}^{(2),v} &=& -\frac{\pi^2}{9}\big(3H_n+5-6\log(2)\big)n^{-1} + \frac{11\sqrt{\pi}}{9}n^{-3/2} + \Order(n^{-2}) , \nonumber \\
  C_{F,n}^{(2),v} &=& -\frac{4\sqrt{\pi}}{9}\big(\pi^2-11\big)n^{-3/2} + \frac{\sqrt{\pi}}{54}\big(33\pi^2-344\big)n^{-5/2} + \Order(n^{-7/2}), 
\end{eqnarray}
where $H_n$ is the harmonic number given by 
\begin{equation}
  H_{n} = \sum_{i=1}^{n}\frac{1}{i} 
\end{equation}
for integer arguments or its generalization 
\begin{equation}
  H_{n} = \gamma_E + \psi_0(n+1)
\end{equation}
for non-integer
arguments, where $\gamma_E$ is Euler's constant and $\psi_0$ is the
digamma function. The asymptotic behavior is in both cases given by 
\begin{equation}
  H_n \approx \gamma_E + \log(n) + \Order(n^{-1}) \ .
\end{equation}
\begin{figure}[!t]
  \begin{center}
    \includegraphics[height=9cm]{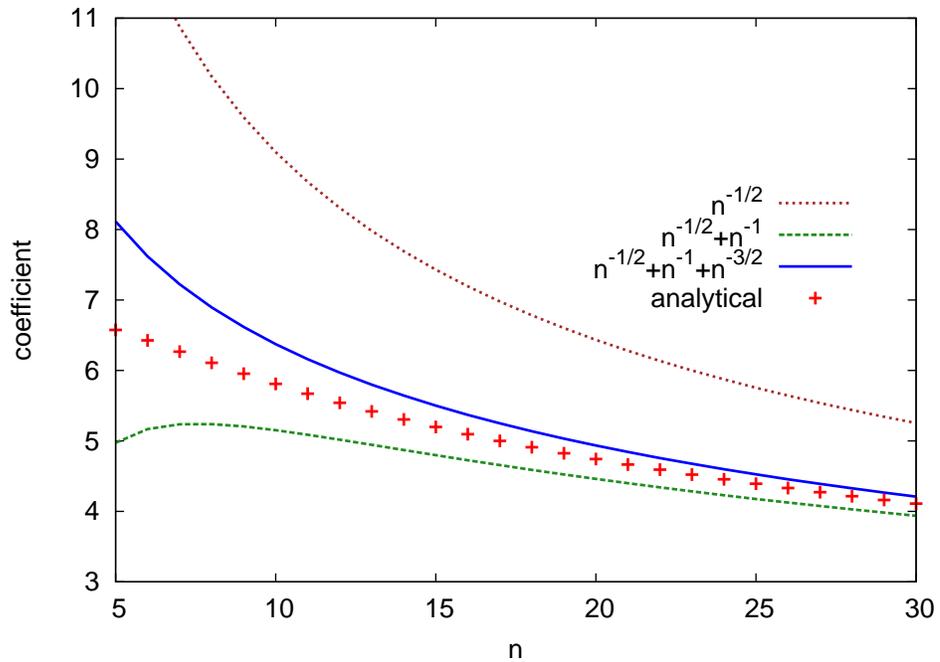}
  \end{center}
  \caption{Analytical moments compared to the large $n$ expansion to different depths for the $C_{A,n}^{(2),v}$ part.\label{largen1}}
  \vspace{5mm}
\end{figure}
\begin{figure}[!t]
  \begin{center}
    \includegraphics[height=9cm]{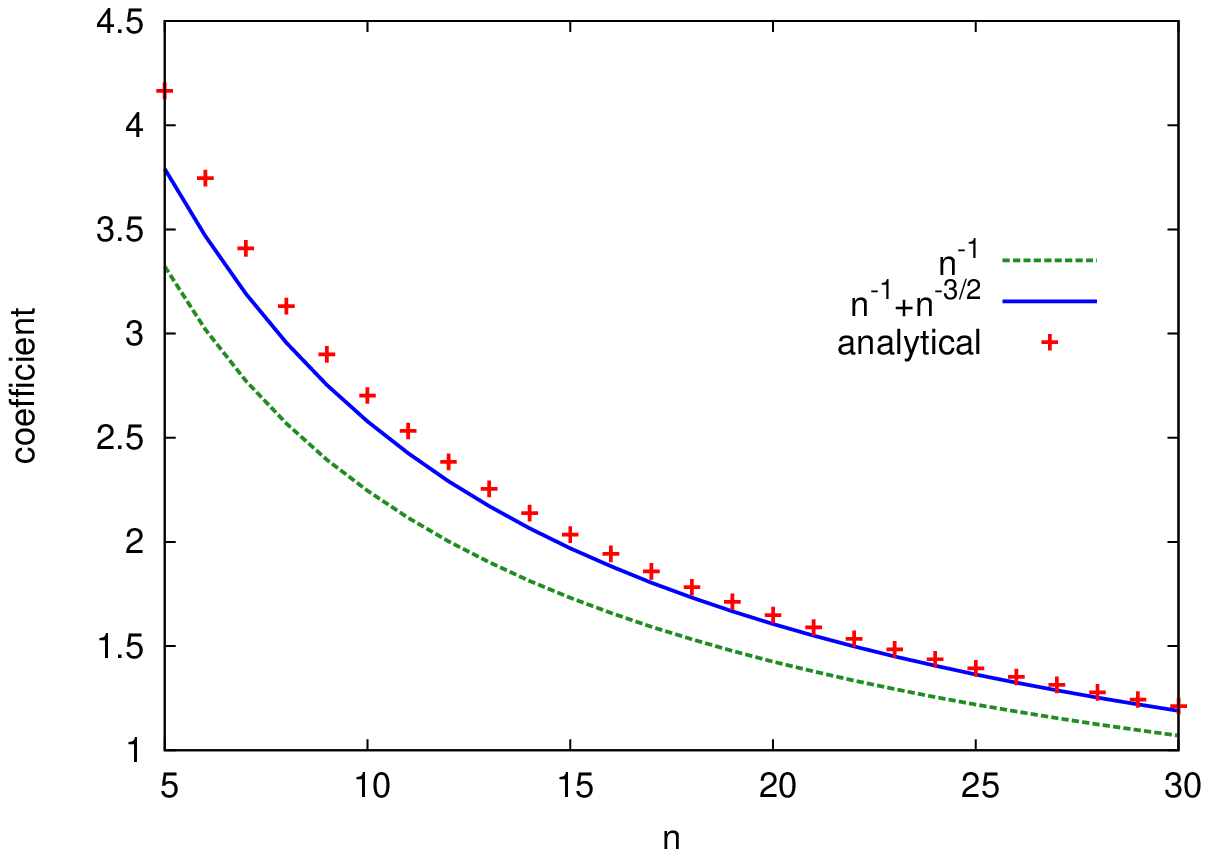}
  \end{center}
  \caption{Analytical moments compared to the large $n$ expansion to different depths for the $C_{NA,n}^{(2),v}$ part.\label{largen2}}
  \vspace{5mm}
\end{figure}
\begin{figure}[!t]
  \begin{center}
    \includegraphics[height=9cm]{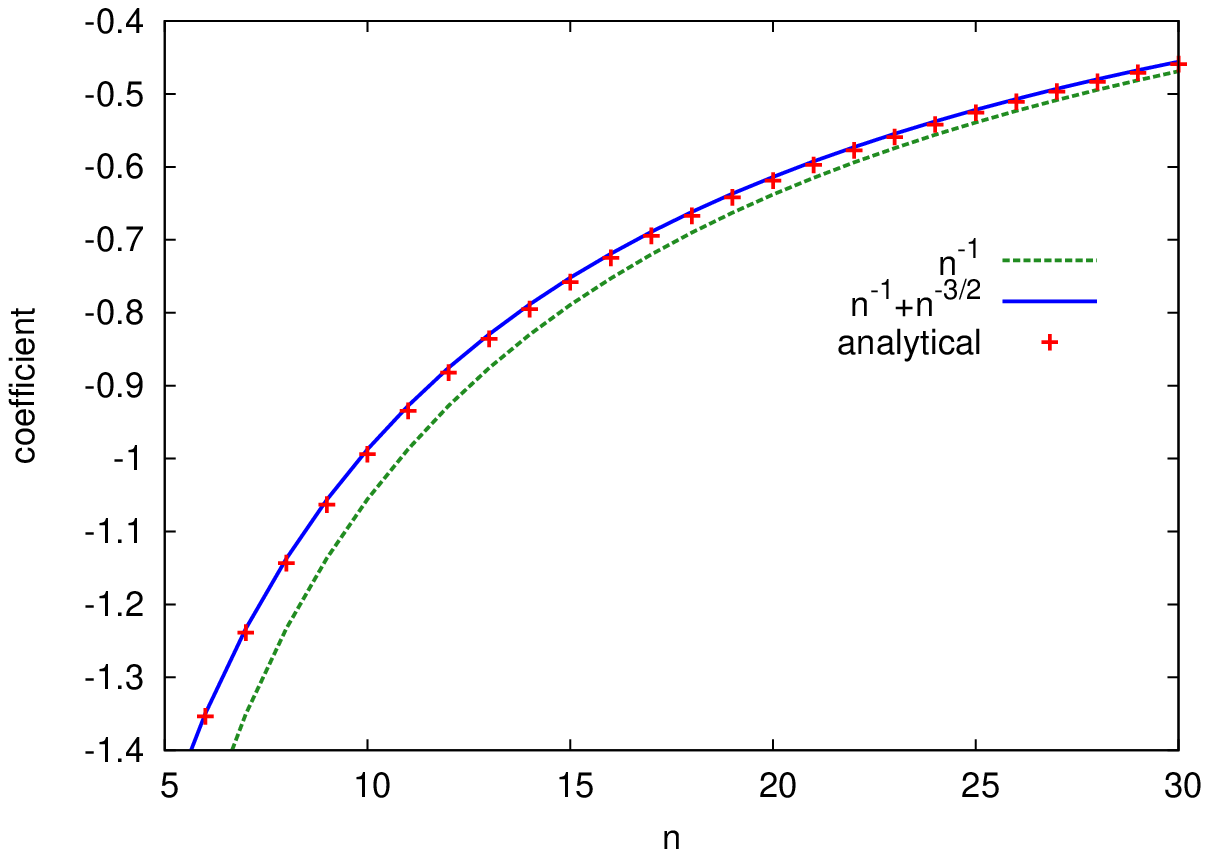}
  \end{center}
  \caption{Analytical moments compared to the large $n$ expansion to different depths for the $C_{l,n}^{(2),v}$ part.\label{largen3}}
  \vspace{5mm}
\end{figure}
\begin{figure}[!t]
  \begin{center}
    \includegraphics[height=9cm]{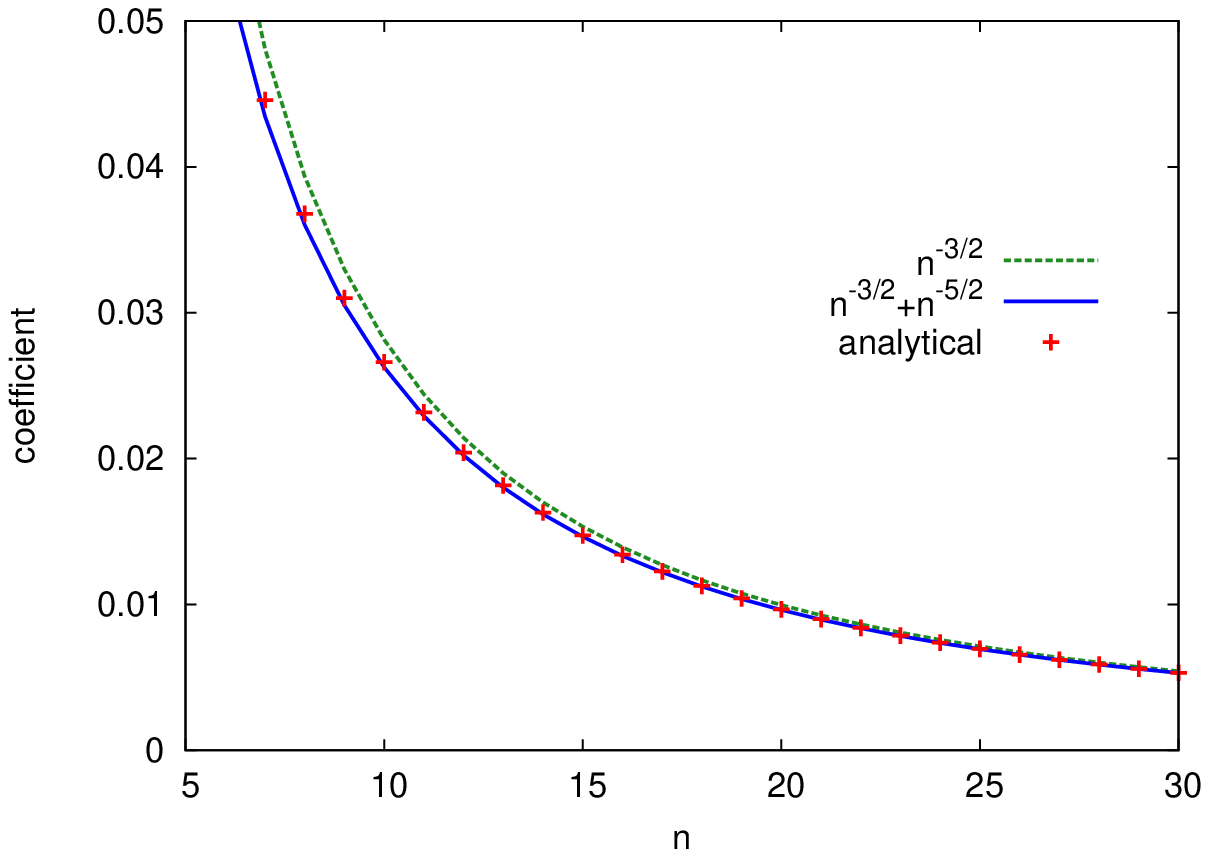}
  \end{center}
  \caption{Analytical moments compared to the large $n$ expansion to different depths for the $C_{F,n}^{(2),v}$ part.\label{largen4}}
  \vspace{5mm}
\end{figure}
\begin{figure}[!t]
  \begin{center}
    \includegraphics[height=9cm]{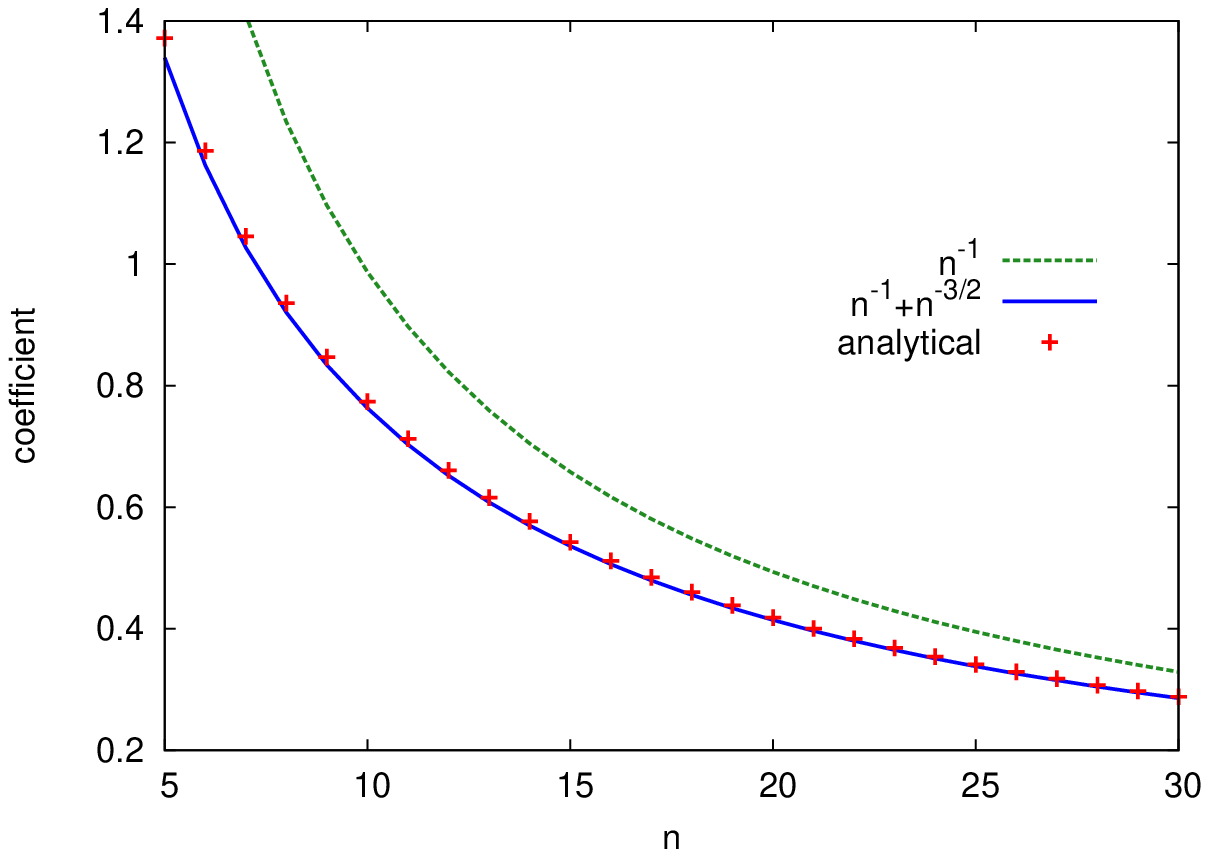}
  \end{center}
  \caption{Analytical moments compared to the large $n$ expansion to different depths for the $C_{n}^{(1),v}$ part.\label{largen5}}
  \vspace{5mm}
\end{figure}

The comparison is shown for the vector current separately for each color
factor in Fig. \ref{largen1}-\ref{largen4}. The
behavior of the leading order approximation is worst for the
$C_{A,n}^{(2),v}$ contribution since the series starts with
$\mathcal{O}(n^{-1/2})$ compared with at least $\mathcal{O}(n^{-1})$ for the
  other color factors. Comparing identical orders in the $n$ expansion
  of the various color structures shows roughly the same degree of
  convergence, but the $C_{A,n}^{(2),v}$ is still penalized by the bad
  leading approximation.  Overall we find good agreement
between the large $n$ expansion and the analytical moments. 
% Apparently,
% for $C_{A,n}^{(2),v}$ the convergence of the expansion is not as good
% as for the others, but as in this case three terms of the expansion are
% known (compared to two for the others), still adequate agreement is
% observed. 
Also at two-loops this procedure shows nearly perfect agreement
if one includes two terms in the large $n$ expansion as can be seen in
Fig. \ref{largen5} .

% ================ %
\section{Conclusion}
% ================ %
\label{sec:conclusion}
We calculated the low-energy expansion of the heavy quark correlator for
the 
scalar, pseudo-scalar, vector and axial-vector current up to $\left (
  \frac{q^2}{4 m^2} \right )^{30}$ in three-loop approximation. For the
scalar, pseudo-scalar and axial-vector currents we included the singlet
contributions arising from double triangle diagrams. Furthermore, for
the vector current we
compared the large $n$ behavior of the moments with  
the asymptotic form derived from threshold behavior. Taking  the lowest
two or three terms of the $1/n$ expansion into account, 
we find good
agreement between the asymptotic formulae and the explicit results.

\section*{Acknowledgments}
\label{sec:ack}
We like to thank J.\,H. K\"uhn, K. Chetyrkin and M. Steinhauser for
fruitful discussions. 
This work was supported by the DFG through SFB/TR~9. The work of Ph.\,M.
was supported by the Graduiertenkolleg ``Hochenergiephysik und
Teilchenastrophysik''. 

\newpage
\appendix

% =========================================== %
\section{Numerical results}\label{sec:app:num}
% =========================================== %
\renewcommand{\arraystretch}{1.3}
\setlength{\tabcolsep}{4pt}
\begin{table}[hptb]
    \centering
  \begin{tabular}[c]{||r|c|c|c|c|c|c||}
    \hline
    n&$C_n^{(0),v}$&$C_n^{(1),v}$&$C_{A,n}^{(2),v}$&$C_{NA,n}^{(2),v}$&$C_{l,n}^{(2),v}$&$C_{F,n}^{(2),v}$\\\hline
1  &   1.067 &     1.916 & 0.1541 & -0.006971 & 0.9934 &    0.3956 \\
2  &  0.4571 &    0.8322 & 0.3025 &    0.2215 & 0.6824 &  -0.01696 \\
3  &  0.2709 &    0.3895 & 0.2711 &  -0.01532 & 0.6433 &   -0.1026 \\
4  &  0.1847 &    0.1523 & 0.2785 &   -0.2218 & 0.6372 &   -0.1301 \\
5  &  0.1364 &  0.007970 & 0.3424 &   -0.3775 & 0.6362 &   -0.1396 \\
6  &  0.1061 &  -0.08688 & 0.4535 &   -0.4936 & 0.6351 &   -0.1421 \\
7  & 0.08558 &   -0.1525 & 0.6007 &   -0.5813 & 0.6329 &   -0.1415 \\
8  & 0.07094 &   -0.1995 & 0.7750 &   -0.6484 & 0.6297 &   -0.1396 \\
9  & 0.06006 &   -0.2342 & 0.9694 &   -0.7004 & 0.6257 &   -0.1369 \\
10 & 0.05170 &   -0.2602 &  1.179 &   -0.7412 & 0.6211 &   -0.1340 \\
11 & 0.04512 &   -0.2801 &  1.400 &   -0.7734 & 0.6162 &   -0.1310 \\
12 & 0.03983 &   -0.2954 &  1.629 &   -0.7991 & 0.6110 &   -0.1281 \\
13 & 0.03550 &   -0.3073 &  1.864 &   -0.8195 & 0.6056 &   -0.1252 \\
14 & 0.03190 &   -0.3166 &  2.104 &   -0.8359 & 0.6002 &   -0.1225 \\
15 & 0.02887 &   -0.3239 &  2.347 &   -0.8490 & 0.5948 &   -0.1198 \\
16 & 0.02629 &   -0.3296 &  2.592 &   -0.8595 & 0.5894 &   -0.1173 \\
17 & 0.02408 &   -0.3339 &  2.838 &   -0.8677 & 0.5841 &   -0.1149 \\
18 & 0.02216 &   -0.3373 &  3.085 &   -0.8742 & 0.5788 &   -0.1126 \\
19 & 0.02048 &   -0.3398 &  3.333 &   -0.8793 & 0.5737 &   -0.1105 \\
20 & 0.01900 &   -0.3417 &  3.581 &   -0.8830 & 0.5686 &   -0.1084 \\
21 & 0.01770 &   -0.3430 &  3.828 &   -0.8858 & 0.5636 &   -0.1065 \\
22 & 0.01653 &   -0.3438 &  4.074 &   -0.8877 & 0.5587 &   -0.1046 \\
23 & 0.01549 &   -0.3443 &  4.319 &   -0.8888 & 0.5540 &   -0.1028 \\
24 & 0.01455 &   -0.3444 &  4.564 &   -0.8894 & 0.5493 &   -0.1011 \\
25 & 0.01371 &   -0.3443 &  4.807 &   -0.8894 & 0.5448 &  -0.09953 \\
26 & 0.01294 &   -0.3439 &  5.049 &   -0.8889 & 0.5404 &  -0.09798 \\
27 & 0.01224 &   -0.3434 &  5.290 &   -0.8881 & 0.5360 &  -0.09650 \\
28 & 0.01161 &   -0.3427 &  5.529 &   -0.8870 & 0.5318 &  -0.09509 \\
29 & 0.01102 &   -0.3419 &  5.767 &   -0.8855 & 0.5277 &  -0.09373 \\
30 & 0.01049 &   -0.3410 &  6.003 &   -0.8839 & 0.5237 &  -0.09243 \\\hline
  \end{tabular}
  \caption{Moments for the vector correlator in the $\overline{\mathrm{MS}}$ scheme}
  \label{tab:vectorMS}
\end{table}
\begin{table}[hptb]
  \centering
  \begin{tabular}[c]{||r|c|c|c|c|c|c||}
    \hline
    n&$C_n^{(0),v}$&$C_n^{(1),v}$&$C_{A,n}^{(2),v}$&$C_{NA,n}^{(2),v}$&$C_{l,n}^{(2),v}$&$C_{F,n}^{(2),v}$\\\hline
1  &   1.067 &  4.049 & 5.075 & 7.098 &  -2.339 &   0.7270 \\
2  &  0.4571 &  2.661 & 6.393 & 6.311 &  -2.174 &   0.2671 \\
3  &  0.2709 &  2.015 & 6.689 & 5.398 &  -1.896 &   0.1499 \\
4  &  0.1847 &  1.630 & 6.685 & 4.699 &  -1.671 &  0.09947 \\
5  &  0.1364 &  1.372 & 6.574 & 4.165 &  -1.494 &  0.07230 \\
6  &  0.1061 &  1.186 & 6.426 & 3.746 &  -1.353 &  0.05566 \\
7  & 0.08558 &  1.046 & 6.267 & 3.409 &  -1.239 &  0.04459 \\
8  & 0.07094 & 0.9356 & 6.108 & 3.132 &  -1.143 &  0.03677 \\
9  & 0.06006 & 0.8469 & 5.955 & 2.900 &  -1.063 &  0.03100 \\
10 & 0.05170 & 0.7738 & 5.809 & 2.702 & -0.9941 &  0.02661 \\
11 & 0.04512 & 0.7126 & 5.671 & 2.532 & -0.9345 &  0.02317 \\
12 & 0.03983 & 0.6605 & 5.542 & 2.384 & -0.8822 &  0.02041 \\
13 & 0.03550 & 0.6156 & 5.420 & 2.254 & -0.8361 &  0.01816 \\
14 & 0.03190 & 0.5765 & 5.305 & 2.139 & -0.7950 &  0.01629 \\
15 & 0.02887 & 0.5422 & 5.198 & 2.035 & -0.7581 &  0.01473 \\
16 & 0.02629 & 0.5118 & 5.096 & 1.943 & -0.7248 &  0.01340 \\
17 & 0.02408 & 0.4847 & 5.000 & 1.858 & -0.6946 &  0.01226 \\
18 & 0.02216 & 0.4603 & 4.910 & 1.782 & -0.6671 &  0.01127 \\
19 & 0.02048 & 0.4383 & 4.824 & 1.712 & -0.6419 &  0.01041 \\
20 & 0.01900 & 0.4184 & 4.743 & 1.648 & -0.6187 & 0.009654 \\
21 & 0.01770 & 0.4002 & 4.665 & 1.589 & -0.5973 & 0.008984 \\
22 & 0.01653 & 0.3836 & 4.592 & 1.535 & -0.5775 & 0.008389 \\
23 & 0.01549 & 0.3683 & 4.522 & 1.484 & -0.5590 & 0.007857 \\
24 & 0.01455 & 0.3541 & 4.455 & 1.437 & -0.5419 & 0.007379 \\
25 & 0.01371 & 0.3411 & 4.392 & 1.393 & -0.5258 & 0.006947 \\
26 & 0.01294 & 0.3290 & 4.331 & 1.352 & -0.5108 & 0.006556 \\
27 & 0.01224 & 0.3177 & 4.272 & 1.314 & -0.4967 & 0.006201 \\
28 & 0.01161 & 0.3072 & 4.216 & 1.277 & -0.4834 & 0.005876 \\
29 & 0.01102 & 0.2974 & 4.162 & 1.243 & -0.4709 & 0.005579 \\
30 & 0.01049 & 0.2881 & 4.111 & 1.211 & -0.4590 & 0.005306 \\\hline
  \end{tabular}
  \caption{Moments for the vector correlator in the onshell scheme}
  \label{tab:vectorOS}
\end{table}
\begin{table}
  \centering
\begin{tabular}{||c|c|c|c|c|c|c|c|c||}\hline
n & $C_n^{(0),a}$ &$ C_n^{(1),a}$ &$ C_{A,n}^{(2),a}$ & $C^{(2),a}_{NA,n}$ &
$ C_{l,n}^{(2),a}$ &$ C_{F,n}^{(2),a}$ &$ C_{S,n}^{(2),a} [1]$ &
$C_{S,n}^{(2),a} [L]$ \\\hline
1  &   0.53330 &    0.6346 & 0.5003 &  -0.8386 &  0.6198 &    0.1775 &     1.970 &    -0.6914 \\
2  &   0.15240 &    0.1062 & 0.3025 &  -0.3370 &  0.2883 &  -0.02960 &    0.3832 &    -0.1867 \\
3  &   0.06772 & -0.009570 & 0.2271 &  -0.2631 &  0.2034 &  -0.04074 &    0.1447 &   -0.08050 \\
4  &   0.03694 &  -0.04310 & 0.2071 &  -0.2263 &  0.1602 &  -0.03578 &   0.07290 &   -0.04309 \\
5  &   0.02273 &  -0.05255 & 0.2079 &  -0.1987 &  0.1323 &  -0.02992 &   0.04308 &   -0.02618 \\
6  &   0.01516 &  -0.05367 & 0.2160 &  -0.1764 &  0.1124 &  -0.02505 &   0.02813 &   -0.01729 \\
7  &   0.01070 &  -0.05172 & 0.2261 &  -0.1578 & 0.09739 &  -0.02121 &   0.01966 &   -0.01212 \\
8  &  0.007883 &  -0.04869 & 0.2363 &  -0.1422 & 0.08563 &  -0.01820 &   0.01443 &  -0.008880 \\
9  &  0.006006 &  -0.04540 & 0.2456 &  -0.1290 & 0.07617 &  -0.01580 &   0.01098 &  -0.006736 \\
10 &  0.004700 &  -0.04219 & 0.2539 &  -0.1177 & 0.06841 &  -0.01388 &  0.008609 &  -0.005252 \\
11 &  0.003760 &  -0.03919 & 0.2610 &  -0.1079 & 0.06193 &  -0.01230 &  0.006905 &  -0.004189 \\
12 &  0.003064 &  -0.03644 & 0.2671 & -0.09949 & 0.05646 &  -0.01099 &  0.005646 &  -0.003404 \\
13 &  0.002536 &  -0.03394 & 0.2723 & -0.09209 & 0.05177 & -0.009902 &  0.004690 &  -0.002811 \\
14 &  0.002127 &  -0.03168 & 0.2767 & -0.08558 & 0.04772 & -0.008976 &  0.003949 &  -0.002353 \\
15 &  0.001804 &  -0.02963 & 0.2803 & -0.07982 & 0.04420 & -0.008185 &  0.003364 &  -0.001993 \\
16 &  0.001547 &  -0.02779 & 0.2833 & -0.07469 & 0.04109 & -0.007502 &  0.002895 &  -0.001706 \\
17 &  0.001338 &  -0.02612 & 0.2858 & -0.07010 & 0.03835 & -0.006909 &  0.002514 &  -0.001473 \\
18 &  0.001166 &  -0.02460 & 0.2879 & -0.06597 & 0.03591 & -0.006389 &  0.002200 &  -0.001283 \\
19 &  0.001024 &  -0.02322 & 0.2895 & -0.06224 & 0.03373 & -0.005931 &  0.001938 &  -0.001125 \\
20 & 0.0009049 &  -0.02196 & 0.2909 & -0.05886 & 0.03176 & -0.005525 &  0.001719 & -0.0009933 \\
21 & 0.0008043 &  -0.02081 & 0.2919 & -0.05578 & 0.02999 & -0.005162 &  0.001533 & -0.0008821 \\
22 & 0.0007188 &  -0.01975 & 0.2926 & -0.05296 & 0.02837 & -0.004838 &  0.001375 & -0.0007876 \\
23 & 0.0006454 &  -0.01878 & 0.2932 & -0.05039 & 0.02691 & -0.004546 &  0.001238 & -0.0007067 \\
24 & 0.0005821 &  -0.01789 & 0.2935 & -0.04802 & 0.02556 & -0.004282 &  0.001120 & -0.0006370 \\
25 & 0.0005272 &  -0.01706 & 0.2937 & -0.04583 & 0.02433 & -0.004043 &  0.001017 & -0.0005765 \\
26 & 0.0004793 &  -0.01630 & 0.2937 & -0.04381 & 0.02320 & -0.003825 & 0.0009275 & -0.0005238 \\
27 & 0.0004372 &  -0.01559 & 0.2936 & -0.04194 & 0.02216 & -0.003626 & 0.0008484 & -0.0004776 \\
28 & 0.0004002 &  -0.01493 & 0.2934 & -0.04020 & 0.02119 & -0.003444 & 0.0007785 & -0.0004369 \\
29 & 0.0003674 &  -0.01432 & 0.2931 & -0.03858 & 0.02029 & -0.003276 & 0.0007164 & -0.0004009 \\
30 & 0.0003382 &  -0.01375 & 0.2927 & -0.03707 & 0.01946 & -0.003121 & 0.0006611 & -0.0003689 \\\hline
\end{tabular}
\caption{Moments for the axial-vector correlator in the $\overline{\mathrm{MS}}$ scheme}
\label{tab:axialMS}
\end{table}
\begin{table}
  \centering
\begin{tabular}{||c|c|c|c|c|c|c|c|c||}\hline
n & $C_n^{(0),a}$ &$ C_n^{(1),a}$ &$ C_{A,n}^{(2),a}$ & $C^{(2),a}_{NA,n}$ &
$ C_{l,n}^{(2),a}$ &$ C_{F,n}^{(2),a}$ &$ C_{S,n}^{(2),a} [1]$ &
$C_{S,n}^{(2),a} [L]$ \\\hline
1  &    0.5333 &    1.701 &   2.314 &   2.714 &   -1.046 &     0.3433 &     1.970 &    -0.6914 \\
2  &    0.1524 &   0.7158 &   1.648 &   1.693 &  -0.6638 &    0.06509 &    0.3832 &    -0.1867 \\
3  &   0.06772 &   0.3968 &   1.190 &   1.090 &  -0.4313 &    0.02238 &    0.1447 &   -0.08050 \\
4  &   0.03694 &   0.2524 &  0.8998 &  0.7579 &  -0.3015 &    0.01013 &   0.07290 &   -0.04309 \\
5  &   0.02273 &   0.1748 &  0.7077 &  0.5583 &  -0.2228 &   0.005396 &   0.04308 &   -0.02618 \\
6  &   0.01516 &   0.1282 &  0.5741 &  0.4293 &  -0.1717 &   0.003202 &   0.02813 &   -0.01729 \\
7  &   0.01070 &  0.09805 &  0.4772 &  0.3409 &  -0.1366 &   0.002052 &   0.01966 &   -0.01212 \\
8  &  0.007883 &  0.07743 &  0.4045 &  0.2778 &  -0.1114 &   0.001393 &   0.01443 &  -0.008880 \\
9  &  0.006006 &  0.06270 &  0.3484 &  0.2310 & -0.09269 &  0.0009894 &   0.01098 &  -0.006736 \\
10 &  0.004700 &  0.05181 &  0.3040 &  0.1954 & -0.07843 &  0.0007283 &  0.008609 &  -0.005252 \\
11 &  0.003760 &  0.04353 &  0.2683 &  0.1675 & -0.06728 &  0.0005520 &  0.006905 &  -0.004189 \\
12 &  0.003064 &  0.03709 &  0.2390 &  0.1454 & -0.05840 &  0.0004287 &  0.005646 &  -0.003404 \\
13 &  0.002536 &  0.03199 &  0.2147 &  0.1275 & -0.05120 &  0.0003399 &  0.004690 &  -0.002811 \\
14 &  0.002127 &  0.02787 &  0.1942 &  0.1127 & -0.04529 &  0.0002742 &  0.003949 &  -0.002353 \\
15 &  0.001804 &  0.02450 &  0.1768 &  0.1005 & -0.04036 &  0.0002246 &  0.003364 &  -0.001993 \\
16 &  0.001547 &  0.02170 &  0.1618 & 0.09013 & -0.03621 &  0.0001864 &  0.002895 &  -0.001706 \\
17 &  0.001338 &  0.01936 &  0.1488 & 0.08136 & -0.03269 &  0.0001566 &  0.002514 &  -0.001473 \\
18 &  0.001166 &  0.01738 &  0.1375 & 0.07384 & -0.02967 &  0.0001329 &  0.002200 &  -0.001283 \\
19 &  0.001024 &  0.01569 &  0.1275 & 0.06734 & -0.02705 &  0.0001138 &  0.001938 &  -0.001125 \\
20 & 0.0009049 &  0.01424 &  0.1187 & 0.06168 & -0.02478 & 0.00009826 &  0.001719 & -0.0009933 \\
21 & 0.0008043 &  0.01297 &  0.1109 & 0.05672 & -0.02278 & 0.00008549 &  0.001533 & -0.0008821 \\
22 & 0.0007188 &  0.01187 &  0.1038 & 0.05236 & -0.02103 & 0.00007488 &  0.001375 & -0.0007876 \\
23 & 0.0006454 &  0.01091 & 0.09753 & 0.04849 & -0.01947 & 0.00006599 &  0.001238 & -0.0007067 \\
24 & 0.0005821 &  0.01005 & 0.09183 & 0.04504 & -0.01808 & 0.00005849 &  0.001120 & -0.0006370 \\
25 & 0.0005272 & 0.009298 & 0.08668 & 0.04196 & -0.01684 & 0.00005211 &  0.001017 & -0.0005765 \\
26 & 0.0004793 & 0.008624 & 0.08198 & 0.03919 & -0.01573 & 0.00004664 & 0.0009275 & -0.0005238 \\
27 & 0.0004372 & 0.008021 & 0.07770 & 0.03669 & -0.01473 & 0.00004194 & 0.0008484 & -0.0004776 \\
28 & 0.0004002 & 0.007479 & 0.07377 & 0.03443 & -0.01382 & 0.00003786 & 0.0007785 & -0.0004369 \\
29 & 0.0003674 & 0.006991 & 0.07017 & 0.03238 & -0.01299 & 0.00003431 & 0.0007164 & -0.0004009 \\
30 & 0.0003382 & 0.006548 & 0.06685 & 0.03051 & -0.01224 & 0.00003120 & 0.0006611 & -0.0003689 \\\hline
\end{tabular}
\caption{Moments for the axial-vector correlator in the onshell  scheme}
\label{tab:axialOS}
\end{table}
\begin{table}
  \centering
\begin{tabular}{||c|c|c|c|c|c|c|c|c||}\hline
n & $C_n^{(0),s}$ &$ C_n^{(1),s}$ &$ C_{A,n}^{(2),s}$ & $C^{(2),s}_{NA,n}$ &
$ C_{l,n}^{(2),s}$ &$ C_{F,n}^{(2),s}$ &$ C_{S,n}^{(2),s} [1]$ &
$C_{S,n}^{(2),s} [L]$ \\\hline
1  &    0.8000 &   0.4519 & 0.03484 &   -2.511 &  0.8815 &    0.7186 &    2.353 &    -0.4444 \\
2  &    0.2286 &   0.3194 &  0.5536 &  -0.6269 &  0.3550 &    0.1201 &   0.8228 &    -0.2074 \\
3  &    0.1016 &   0.1152 &  0.4516 &  -0.3544 &  0.2345 &  0.008315 &   0.3660 &    -0.1089 \\
4  &   0.05541 &  0.02460 &  0.3448 &  -0.2797 &  0.1857 &  -0.01803 &   0.1923 &   -0.06377 \\
5  &   0.03410 & -0.01630 &  0.2876 &  -0.2439 &  0.1570 &  -0.02406 &   0.1134 &   -0.04057 \\
6  &   0.02273 & -0.03517 &  0.2635 &  -0.2193 &  0.1366 &  -0.02428 &  0.07267 &   -0.02746 \\
7  &   0.01605 & -0.04364 &  0.2577 &  -0.1995 &  0.1210 &  -0.02277 &  0.04956 &   -0.01950 \\
8  &   0.01182 & -0.04691 &  0.2614 &  -0.1827 &  0.1083 &  -0.02083 &  0.03546 &   -0.01438 \\
9  &  0.009009 & -0.04752 &  0.2698 &  -0.1681 & 0.09788 &  -0.01890 &  0.02635 &   -0.01093 \\
10 &  0.007050 & -0.04673 &  0.2803 &  -0.1553 & 0.08908 &  -0.01713 &  0.02019 &  -0.008525 \\
11 &  0.005640 & -0.04524 &  0.2913 &  -0.1440 & 0.08156 &  -0.01556 &  0.01586 &  -0.006789 \\
12 &  0.004596 & -0.04342 &  0.3021 &  -0.1339 & 0.07508 &  -0.01417 &  0.01273 &  -0.005505 \\
13 &  0.003803 & -0.04146 &  0.3124 &  -0.1250 & 0.06943 &  -0.01296 &  0.01040 &  -0.004532 \\
14 &  0.003190 & -0.03949 &  0.3219 &  -0.1170 & 0.06448 &  -0.01190 & 0.008624 &  -0.003782 \\
15 &  0.002707 & -0.03756 &  0.3305 &  -0.1098 & 0.06010 &  -0.01096 & 0.007246 &  -0.003193 \\
16 &  0.002320 & -0.03571 &  0.3383 &  -0.1033 & 0.05621 &  -0.01014 & 0.006159 &  -0.002723 \\
17 &  0.002006 & -0.03397 &  0.3454 & -0.09739 & 0.05273 & -0.009409 & 0.005287 &  -0.002344 \\
18 &  0.001749 & -0.03232 &  0.3516 & -0.09205 & 0.04960 & -0.008759 & 0.004580 &  -0.002034 \\
19 &  0.001536 & -0.03078 &  0.3572 & -0.08719 & 0.04678 & -0.008178 & 0.003998 &  -0.001778 \\
20 &  0.001357 & -0.02934 &  0.3622 & -0.08274 & 0.04422 & -0.007656 & 0.003516 &  -0.001565 \\
21 &  0.001206 & -0.02799 &  0.3665 & -0.07866 & 0.04189 & -0.007187 & 0.003111 &  -0.001385 \\
22 &  0.001078 & -0.02674 &  0.3704 & -0.07491 & 0.03976 & -0.006763 & 0.002769 &  -0.001233 \\
23 & 0.0009681 & -0.02556 &  0.3738 & -0.07145 & 0.03781 & -0.006378 & 0.002478 &  -0.001103 \\
24 & 0.0008732 & -0.02447 &  0.3767 & -0.06826 & 0.03602 & -0.006027 & 0.002228 & -0.0009917 \\
25 & 0.0007908 & -0.02344 &  0.3793 & -0.06530 & 0.03437 & -0.005707 & 0.002012 & -0.0008952 \\
26 & 0.0007189 & -0.02249 &  0.3816 & -0.06255 & 0.03285 & -0.005415 & 0.001825 & -0.0008113 \\
27 & 0.0006559 & -0.02159 &  0.3835 & -0.05999 & 0.03144 & -0.005146 & 0.001661 & -0.0007379 \\
28 & 0.0006003 & -0.02075 &  0.3852 & -0.05761 & 0.03012 & -0.004898 & 0.001517 & -0.0006734 \\
29 & 0.0005511 & -0.01996 &  0.3866 & -0.05538 & 0.02890 & -0.004669 & 0.001390 & -0.0006165 \\
30 & 0.0005073 & -0.01922 &  0.3878 & -0.05329 & 0.02776 & -0.004458 & 0.001278 & -0.0005661 \\\hline
\end{tabular}
\caption{Moments for the scalar correlator in the $\overline{\mathrm{MS}}$  scheme}
\label{tab:scalarMS}
\end{table}
\begin{table}
  \centering
\begin{tabular}{||c|c|c|c|c|c|c|c|c||}\hline
n & $C_n^{(0),s}$ &$ C_n^{(1),s}$ &$ C_{A,n}^{(2),s}$ & $C^{(2),s}_{NA,n}$ &
$ C_{l,n}^{(2),s}$ &$ C_{F,n}^{(2),s}$ &$ C_{S,n}^{(2),s} [1]$ &
$C_{S,n}^{(2),s} [L]$ \\\hline
1  &    0.8000 &  0.4519 & 0.03484 &  -2.511 &   0.8815 &    0.7186 &    2.353 &    -0.4444 \\
2  &    0.2286 &  0.7765 &   1.426 &  0.8955 &  -0.3591 &    0.1911 &   0.8228 &    -0.2074 \\
3  &    0.1016 &  0.5215 &   1.526 &  0.9989 &  -0.4002 &   0.07144 &   0.3660 &    -0.1089 \\
4  &   0.05541 &  0.3571 &   1.327 &  0.8275 &  -0.3336 &   0.03362 &   0.1923 &   -0.06377 \\
5  &   0.03410 &  0.2565 &   1.115 &  0.6646 &  -0.2691 &   0.01832 &   0.1134 &   -0.04057 \\
6  &   0.02273 &  0.1922 &  0.9371 &  0.5378 &  -0.2185 &   0.01103 &  0.07267 &   -0.02746 \\
7  &   0.01605 &  0.1489 &  0.7952 &  0.4417 &  -0.1798 &  0.007146 &  0.04956 &   -0.01950 \\
8  &   0.01182 &  0.1186 &  0.6824 &  0.3685 &  -0.1502 &  0.004890 &  0.03546 &   -0.01438 \\
9  &  0.009009 & 0.09662 &  0.5921 &  0.3119 &  -0.1273 &  0.003494 &  0.02635 &   -0.01093 \\
10 &  0.007050 & 0.08017 &  0.5191 &  0.2673 &  -0.1092 &  0.002584 &  0.02019 &  -0.008525 \\
11 &  0.005640 & 0.06756 &  0.4592 &  0.2317 & -0.09464 &  0.001967 &  0.01586 &  -0.006789 \\
12 &  0.004596 & 0.05769 &  0.4096 &  0.2028 & -0.08285 &  0.001532 &  0.01273 &  -0.005505 \\
13 &  0.003803 & 0.04982 &  0.3681 &  0.1790 & -0.07315 &  0.001218 &  0.01040 &  -0.004532 \\
14 &  0.003190 & 0.04345 &  0.3329 &  0.1592 & -0.06508 & 0.0009850 & 0.008624 &  -0.003782 \\
15 &  0.002707 & 0.03823 &  0.3028 &  0.1426 & -0.05828 & 0.0008084 & 0.007246 &  -0.003193 \\
16 &  0.002320 & 0.03388 &  0.2769 &  0.1285 & -0.05251 & 0.0006722 & 0.006159 &  -0.002723 \\
17 &  0.002006 & 0.03024 &  0.2544 &  0.1164 & -0.04757 & 0.0005654 & 0.005287 &  -0.002344 \\
18 &  0.001749 & 0.02715 &  0.2347 &  0.1060 & -0.04330 & 0.0004804 & 0.004580 &  -0.002034 \\
19 &  0.001536 & 0.02451 &  0.2174 & 0.09695 & -0.03959 & 0.0004119 & 0.003998 &  -0.001778 \\
20 &  0.001357 & 0.02224 &  0.2020 & 0.08903 & -0.03635 & 0.0003561 & 0.003516 &  -0.001565 \\
21 &  0.001206 & 0.02027 &  0.1884 & 0.08206 & -0.03349 & 0.0003101 & 0.003111 &  -0.001385 \\
22 &  0.001078 & 0.01855 &  0.1762 & 0.07589 & -0.03097 & 0.0002719 & 0.002769 &  -0.001233 \\
23 & 0.0009681 & 0.01703 &  0.1653 & 0.07041 & -0.02872 & 0.0002398 & 0.002478 &  -0.001103 \\
24 & 0.0008732 & 0.01570 &  0.1554 & 0.06551 & -0.02672 & 0.0002127 & 0.002228 & -0.0009917 \\
25 & 0.0007908 & 0.01451 &  0.1464 & 0.06112 & -0.02492 & 0.0001896 & 0.002012 & -0.0008952 \\
26 & 0.0007189 & 0.01346 &  0.1383 & 0.05716 & -0.02330 & 0.0001698 & 0.001825 & -0.0008113 \\
27 & 0.0006559 & 0.01252 &  0.1309 & 0.05359 & -0.02184 & 0.0001528 & 0.001661 & -0.0007379 \\
28 & 0.0006003 & 0.01167 &  0.1241 & 0.05035 & -0.02051 & 0.0001380 & 0.001517 & -0.0006734 \\
29 & 0.0005511 & 0.01090 &  0.1179 & 0.04739 & -0.01930 & 0.0001251 & 0.001390 & -0.0006165 \\
30 & 0.0005073 & 0.01021 &  0.1122 & 0.04470 & -0.01820 & 0.0001138 & 0.001278 & -0.0005661 \\\hline
\end{tabular}
\caption{Moments for the scalar correlator in the onshell scheme}
\label{tab:scalarOS}
\end{table}
\begin{table}
  \centering
\begin{tabular}{||c|c|c|c|c|c|c|c|c||}\hline
n & $C_n^{(0),p}$ &$ C_n^{(1),p}$ &$ C_{A,n}^{(2),p}$ & $C^{(2),p}_{NA,n}$ &
$ C_{l,n}^{(2),p}$ &$ C_{F,n}^{(2),p}$ &$ C_{S,n}^{(2),p} [1]$ &
$C_{S,n}^{(2),p} [L]$ \\\hline
1  &   1.333 &    2.333 & 2.712 &  -1.858 & 0.9259 &    1.311 &   5.629 &   -1.000 \\
2  &  0.5333 &    1.548 & 3.399 & 0.02615 & 0.4346 &   0.3307 &   3.065 &  -0.6667 \\
3  &  0.3048 &   0.9088 & 2.797 & 0.04435 & 0.4017 &  0.06951 &   1.952 &  -0.4667 \\
4  &  0.2032 &   0.5346 & 2.198 & -0.1204 & 0.4294 & -0.02758 &   1.375 &  -0.3471 \\
5  &  0.1478 &   0.3010 & 1.774 & -0.2777 & 0.4600 & -0.07127 &   1.034 &  -0.2703 \\
6  &  0.1137 &   0.1458 & 1.506 & -0.4046 & 0.4846 & -0.09305 &  0.8132 &  -0.2180 \\
7  & 0.09093 &  0.03751 & 1.358 & -0.5039 & 0.5029 &  -0.1044 &  0.6612 &  -0.1804 \\
8  & 0.07488 & -0.04085 & 1.300 & -0.5813 & 0.5162 &  -0.1104 &  0.5512 &  -0.1525 \\
9  & 0.06306 & -0.09923 & 1.310 & -0.6422 & 0.5256 &  -0.1133 &  0.4686 &  -0.1311 \\
10 & 0.05405 &  -0.1437 & 1.369 & -0.6903 & 0.5320 &  -0.1144 &  0.4048 &  -0.1142 \\
11 & 0.04700 &  -0.1783 & 1.467 & -0.7286 & 0.5362 &  -0.1145 &  0.3542 &  -0.1007 \\
12 & 0.04136 &  -0.2055 & 1.595 & -0.7594 & 0.5386 &  -0.1139 &  0.3134 & -0.08963 \\
13 & 0.03677 &  -0.2271 & 1.745 & -0.7842 & 0.5398 &  -0.1128 &  0.2798 & -0.08044 \\
14 & 0.03296 &  -0.2446 & 1.913 & -0.8042 & 0.5400 &  -0.1116 &  0.2519 & -0.07273 \\
15 & 0.02977 &  -0.2587 & 2.095 & -0.8205 & 0.5394 &  -0.1101 &  0.2283 & -0.06617 \\
16 & 0.02707 &  -0.2702 & 2.288 & -0.8336 & 0.5383 &  -0.1086 &  0.2082 & -0.06054 \\
17 & 0.02475 &  -0.2796 & 2.490 & -0.8442 & 0.5367 &  -0.1071 &  0.1908 & -0.05566 \\
18 & 0.02274 &  -0.2874 & 2.699 & -0.8527 & 0.5347 &  -0.1056 &  0.1758 & -0.05141 \\
19 & 0.02099 &  -0.2937 & 2.913 & -0.8595 & 0.5324 &  -0.1040 &  0.1626 & -0.04767 \\
20 & 0.01945 &  -0.2989 & 3.131 & -0.8649 & 0.5299 &  -0.1025 &  0.1510 & -0.04436 \\
21 & 0.01810 &  -0.3032 & 3.353 & -0.8690 & 0.5273 &  -0.1010 &  0.1407 & -0.04142 \\
22 & 0.01689 &  -0.3067 & 3.577 & -0.8721 & 0.5246 & -0.09956 &  0.1315 & -0.03879 \\
23 & 0.01581 &  -0.3095 & 3.803 & -0.8744 & 0.5217 & -0.09814 &  0.1233 & -0.03642 \\
24 & 0.01484 &  -0.3118 & 4.030 & -0.8759 & 0.5188 & -0.09677 &  0.1159 & -0.03429 \\
25 & 0.01397 &  -0.3136 & 4.258 & -0.8768 & 0.5159 & -0.09544 &  0.1092 & -0.03236 \\
26 & 0.01318 &  -0.3150 & 4.487 & -0.8771 & 0.5129 & -0.09415 &  0.1032 & -0.03060 \\
27 & 0.01246 &  -0.3160 & 4.715 & -0.8770 & 0.5099 & -0.09291 & 0.09765 & -0.02899 \\
28 & 0.01181 &  -0.3168 & 4.944 & -0.8765 & 0.5069 & -0.09170 & 0.09260 & -0.02752 \\
29 & 0.01121 &  -0.3173 & 5.173 & -0.8757 & 0.5039 & -0.09053 & 0.08797 & -0.02618 \\
30 & 0.01065 &  -0.3176 & 5.401 & -0.8746 & 0.5009 & -0.08940 & 0.08372 & -0.02493 \\\hline
\end{tabular}
\caption{Moments for the pseudo-scalar correlator in the $\overline{\mathrm{MS}}$  scheme}
\label{tab:pseudoMS}
\end{table}
\begin{table}
  \centering
\begin{tabular}{||c|c|c|c|c|c|c|c|c||}\hline
n & $C_n^{(0),p}$ &$ C_n^{(1),p}$ &$ C_{A,n}^{(2),p}$ & $C^{(2),p}_{NA,n}$ &
$ C_{l,n}^{(2),p}$ &$ C_{F,n}^{(2),p}$ &$ C_{S,n}^{(2),p} [1]$ &
$C_{S,n}^{(2),p} [L]$ \\\hline
1  &   1.333 &  2.333 & 2.712 & -1.858 &  0.9259 &    1.311 &   5.629 &   -1.000 \\
2  &  0.5333 &  2.615 & 7.040 &  3.578 &  -1.232 &   0.4964 &   3.065 &  -0.6667 \\
3  &  0.3048 &  2.128 & 8.273 &  4.104 &  -1.502 &   0.2589 &   1.952 &  -0.4667 \\
4  &  0.2032 &  1.754 & 8.466 &  3.939 &  -1.475 &   0.1618 &   1.375 &  -0.3471 \\
5  &  0.1478 &  1.483 & 8.331 &  3.659 &  -1.387 &   0.1124 &   1.034 &  -0.2703 \\
6  &  0.1137 &  1.282 & 8.090 &  3.381 &  -1.291 &  0.08353 &  0.8132 &  -0.2180 \\
7  & 0.09093 &  1.129 & 7.821 &  3.130 &  -1.202 &  0.06509 &  0.6612 &  -0.1804 \\
8  & 0.07488 &  1.008 & 7.554 &  2.910 &  -1.121 &  0.05249 &  0.5512 &  -0.1525 \\
9  & 0.06306 & 0.9097 & 7.300 &  2.718 &  -1.050 &  0.04347 &  0.4686 &  -0.1311 \\
10 & 0.05405 & 0.8292 & 7.062 &  2.550 & -0.9878 &  0.03674 &  0.4048 &  -0.1142 \\
11 & 0.04700 & 0.7618 & 6.842 &  2.402 & -0.9322 &  0.03157 &  0.3542 &  -0.1007 \\
12 & 0.04136 & 0.7045 & 6.639 &  2.271 & -0.8828 &  0.02751 &  0.3134 & -0.08963 \\
13 & 0.03677 & 0.6553 & 6.451 &  2.154 & -0.8385 &  0.02424 &  0.2798 & -0.08044 \\
14 & 0.03296 & 0.6125 & 6.276 &  2.050 & -0.7987 &  0.02157 &  0.2519 & -0.07273 \\
15 & 0.02977 & 0.5749 & 6.115 &  1.956 & -0.7627 &  0.01936 &  0.2283 & -0.06617 \\
16 & 0.02707 & 0.5418 & 5.965 &  1.871 & -0.7301 &  0.01749 &  0.2082 & -0.06054 \\
17 & 0.02475 & 0.5122 & 5.825 &  1.793 & -0.7003 &  0.01591 &  0.1908 & -0.05566 \\
18 & 0.02274 & 0.4858 & 5.694 &  1.722 & -0.6730 &  0.01455 &  0.1758 & -0.05141 \\
19 & 0.02099 & 0.4619 & 5.571 &  1.657 & -0.6480 &  0.01338 &  0.1626 & -0.04767 \\
20 & 0.01945 & 0.4403 & 5.456 &  1.597 & -0.6248 &  0.01235 &  0.1510 & -0.04436 \\
21 & 0.01810 & 0.4207 & 5.348 &  1.542 & -0.6034 &  0.01145 &  0.1407 & -0.04142 \\
22 & 0.01689 & 0.4027 & 5.246 &  1.490 & -0.5836 &  0.01065 &  0.1315 & -0.03879 \\
23 & 0.01581 & 0.3862 & 5.150 &  1.443 & -0.5651 & 0.009943 &  0.1233 & -0.03642 \\
24 & 0.01484 & 0.3710 & 5.059 &  1.398 & -0.5478 & 0.009309 &  0.1159 & -0.03429 \\
25 & 0.01397 & 0.3570 & 4.972 &  1.357 & -0.5317 & 0.008739 &  0.1092 & -0.03236 \\
26 & 0.01318 & 0.3440 & 4.890 &  1.318 & -0.5166 & 0.008225 &  0.1032 & -0.03060 \\
27 & 0.01246 & 0.3320 & 4.812 &  1.281 & -0.5023 & 0.007760 & 0.09765 & -0.02899 \\
28 & 0.01181 & 0.3207 & 4.737 &  1.247 & -0.4889 & 0.007337 & 0.09260 & -0.02752 \\
29 & 0.01121 & 0.3102 & 4.666 &  1.214 & -0.4763 & 0.006951 & 0.08797 & -0.02618 \\
30 & 0.01065 & 0.3004 & 4.598 &  1.183 & -0.4643 & 0.006597 & 0.08372 & -0.02493 \\\hline
\end{tabular}
\caption{Moments for the pseudo-scalar correlator in the onshell  scheme}
\label{tab:pseudoOS}
\end{table}

\clearpage

% =========================================== %
\section{Analytical Results}\label{sec:app:ana}
% =========================================== %

In this appendix we present the analytical results for the moments 9 to
12 of the various currents in the $\overline{\mathrm{MS}}$-scheme. The
expression for the moments up to
30 can be obtained from\\
\texttt{http://www-ttp.particle.uni-karlsruhe.de/Progdata/ttp07/ttp07-32}\\
in computer readable form in the $\overline{\mathrm{MS}}$ and onshell
scheme. In the following expressions we use $L=\log \mu^2 / m^2$ and $L_{-q^2}
= \log(-q^2/m^2)$. The singlet contributions
$C_{S,n}^{\left(2\right),\delta}$ are always given separately.
 { \allowdisplaybreaks

\subsection{Scalar Current}
%\label{sec:app:scalar}
% scalar 1loop and 2loop
\begin{align}
  \sum_{n=9}^{12} C_n^{\left(0\right),s} z^n = \frac{131072}{14549535}z^9 +
  \frac{262144}{37182145}z^{10} + \frac{1048576}{185910725}z^{11}+ \frac{2097152}{456326325}z^{12}
\nonumber
\end{align}

\begin{align}
  \sum_{n=9}^{12} C_n^{(1),s} z^n =& 
  \left(
    - \frac{4979043746471936}{104786039514531375}
    - \frac{524288}{4849845}L
  \right)z^9\noeqnumlb
  &+ \left(
    - \frac{2888036521295872}{61796895098313375}
    - \frac{3538944}{37182145}L
  \right)z^{10}\noeqnumlb
  &+ \left(
    - \frac{38132976301932544}{842822916656645925}
    - \frac{3145728}{37182145}L
  \right)z^{11}\noeqnumlb
  &+ \left(
    - \frac{33183292389464276992}{764287144877276645625}
    - \frac{11534336}{152108775}L
  \right)z^{12}
  \nonumber \end{align}

\begin{align}
%  \label{tlsc:threeloopsc009}
&C_{9}^{\left(2\right),s}=\bigg(\frac{1048576}{1616615} L^2+\frac{25108082269650944}{34928679838177125} L\noeqnumlb 
&+\frac{73746539708944133873833421163704138849}{5771391802019474808577720320000}
-\frac{3409922386710494034767011}{320780855251304448} \zeta_3 \bigg) C_F^2 \noeqnumlb 
&+C_A \bigg(-\frac{65536}{1322685} L^2-\frac{5406868428560384}{28578010776690375} L\noeqnumlb
&+\frac{102915497024490763942656449854922612500879}{437529212511096385238276977459200000}-\frac{1345087895588279678743561}{6873875469670809600} \zeta_3 \bigg) C_F\noeqnumlb
&+T_F \bigg(\bigg(\frac{262144}{14549535} n_h+\frac{262144}{14549535} n_l\bigg) L^2+\bigg(\frac{14418875171495936}{314358118543594125} n_h\noeqnumlb
&+\frac{14418875171495936}{314358118543594125} n_l\bigg) L-\frac{370939064258803430677307}{488861314420492468224000} n_h\noeqnumlb
&+\frac{12790934874367656891392}{130678984236690621356625} n_l+\frac{125624633262191}{204096845905920} n_h \zeta_3 \bigg) C_F
\nonumber
\end{align}

\begin{align}
%  \label{tlsc:threeloopsc010}
&C_{10}^{\left(2\right),s}=\bigg(\frac{23887872}{37182145} L^2+\frac{1743551373172736}{2288773892530125} L\noeqnumlb
&+\frac{1181283682867330857434847115794497652295927361}{18741992087247811295709473066188800000}\noeqnumlb
&-\frac{1856910365195457777240839095367}{35414206419744011059200} \zeta_3 \bigg) C_F^2 \noeqnumlb 
&+C_A \bigg(-\frac{147456}{3380195} L^2-\frac{2883103404986368}{16853698663176375} L\noeqnumlb
&+\frac{19485093544688649029885182650128118344232413}{20183683786266873703071740225126400000} \noeqnumlb
&-\frac{738743868672823822714945309}{919849517395948339200} \zeta_3 \bigg) C_F\noeqnumlb
&+T_F \bigg(\bigg(\frac{589824}{37182145} n_h+\frac{589824}{37182145} n_l\bigg) L^2+\bigg(\frac{7789487453519872}{185390685294940125} n_h\noeqnumlb
&+\frac{7789487453519872}{185390685294940125} n_l\bigg) L-\frac{4979656088382493508774203867}{6626352163198301909286912000} n_h\noeqnumlb
&+\frac{48054627678191205275648}{539469652874543334318375} n_l+\frac{332499096340061}{544258255749120} n_h \zeta_3 \bigg) C_F
\nonumber
\end{align}

\begin{align}
%  \label{tlsc:threeloopsc011}
&C_{11}^{\left(2\right),s}=\bigg(\frac{4718592}{7436429} L^2+\frac{8933862862782464}{11237638888755279} L\noeqnumlb
&+\frac{7404906335674218474752347314237559396292812077443}{24289621745073163439239477093780684800000}\noeqnumlb
&-\frac{61139862184252373166516579786803}{241074161519130140737536} \zeta_3 \bigg) C_F^2+C_A \bigg(-\frac{131072}{3380195} L^2\noeqnumlb
&-\frac{1880663970832384}{12097936602726975} L\noeqnumlb
&+\frac{8830293913654550377524203305197260449225292386979}{2237852825236511033113012799650733752320000}\noeqnumlb
&-\frac{209352182103823255038597110761}{63776233206119084851200} \zeta_3 \bigg) C_F+T_F \bigg(\bigg(\frac{524288}{37182145} n_h+\frac{524288}{37182145} n_l\bigg) L^2\noeqnumlb
&+\bigg(\frac{97554230956294144}{2528468749969937775} n_h+\frac{97554230956294144}{2528468749969937775} n_l\bigg) L\noeqnumlb
&-\frac{202040916959120787173924674741}{271479640140730429738057728000} n_h+\frac{394357158107030025924608}{4835000144023764273824265} n_l\noeqnumlb
&+\frac{2287436936895761}{3773523906527232} n_h \zeta_3 \bigg) C_F
\nonumber
\end{align}

\begin{align}
%  \label{tlsc:threeloopsc012}
&C_{12}^{\left(2\right),s}=\bigg(\frac{31719424}{50702925} L^2+\frac{19006459960269111296}{23160216511432625625} L\noeqnumlb
&+\frac{188473078860503880648883122776543550015991250654673059}{129919789020675223866838798646389820620800000}\noeqnumlb
&-\frac{23090275323409880205366729096347}{19132869961835725455360} \zeta_3 \bigg) C_F^2+C_A \bigg(-\frac{15859712}{456326325} L^2\noeqnumlb
&-\frac{29590089938919374848}{208441948602893630625} L\noeqnumlb
&+\frac{20886533237898744255234695965842642441637712801424593}{1299197890206752238668387986463898206208000000}\noeqnumlb
&-\frac{41752924406364782322130747883}{3121913513586248908800} \zeta_3 \bigg) C_F+T_F \bigg(\bigg(\frac{5767168}{456326325} n_h+\frac{5767168}{456326325} n_l\bigg) L^2\noeqnumlb
&+\bigg(\frac{81479567700203732992}{2292861434631829936875} n_h+\frac{81479567700203732992}{2292861434631829936875} n_l\bigg) L\noeqnumlb
&-\frac{88884847785166765220997026873}{120599609370209094749021798400} n_h+\frac{41147533572114438909275987968}{548058254961784643538602765625} n_l\noeqnumlb
&+\frac{77358608492779}{128642860449792} n_h \zeta_3 \bigg) C_F
\nonumber
\end{align}

\begin{align}
  C_{S,9}^{\left(2\right),s} = - \frac{28771023505735526373476647}{1454362410400965092966400000}
  + \frac{143600001918689}{3741775508275200}\zeta_3
  - \frac{289369094656}{26467422856875}L_{-q^2}
  \nonumber \end{align}

\begin{align}
  C_{S,10}^{\left(2\right),s} =
  - \frac{1090589835840240580931001947}{57980581427985141706260480000}
  + \frac{48559274491381}{1496710203310080}\zeta_3
  - \frac{3471392768}{407191120875}L_{-q^2}
  \nonumber \end{align}

\begin{align}
  C_{S,11}^{\left(2\right),s} =&
  - \frac{115469790307862708811970912146409}{
    6584738671613416573296590192640000}
  + \frac{4337059242964423}{156086670679080960}\zeta_3\noeqnumlb
  &- \frac{225667578742784}{33237789623663625}L_{-q^2}
  \nonumber \end{align}
\begin{align}
  C_{S,12}^{\left(2\right),s} =&
  - \frac{6397059251470658815951587545516867}{395084320296804994397795411558400000}
  + \frac{675994471643152871}{28095600722234572800}\zeta_3\noeqnumlb
  &- \frac{2744536517378048}{498566844354954375}L_{-q^2}
  \nonumber \end{align}

\subsection{Pseudo-Scalar Current}
%\label{sec:app:ps}
% pseudoscalar 1loop and 2loop
%
\begin{align}
  \sum_{n=9}^{12} C_n^{\left(0\right),p} z^n =  + \frac{131072}{2078505} z^9 +
  \frac{262144}{4849845} z^{10} + \frac{1048576}{22309287} z^{11} +
  \frac{2097152}{50702925} z^{12}
\nonumber
\end{align}

\begin{align}
  \sum_{n=9}^{12} C_n^{\left(1\right),p} z^n =& 
  \left(
    - \frac{49751393148928}{501368610117375}
    - \frac{524288}{692835}L
  \right)z^9\noeqnumlb
  &+   \left(
    - \frac{557796221206528}{3880964426464125}
    - \frac{1179648}{1616615}L
  \right)z^{10}\noeqnumlb
  &+   \left(
    - \frac{4091600265773056}{22953132465087825}
    - \frac{5242880}{7436429}L
  \right)z^{11}\noeqnumlb
  &+   \left(
    - \frac{35891639960928256}{174694204543377519}
    - \frac{11534336}{16900975}L
  \right)z^{12}
  \nonumber \end{align}

\begin{align}
  C_{9}^{(2),p} =&+ C_FT_R  \bigg(
  \frac{24213861812657343488}{46071664047183800025}n_l
  - \frac{23936615221766899739377}{13951233275688124416000}n_h\noeqnumlb
  & + \frac{386477922523}{289910292480}n_h\zeta_3
  + \frac{365917995900928}{1504105830352125}L(n_l+n_h)
  + \frac{262144}{2078505}L^2(n_l+n_h)
  \bigg)\noeqnumlb
  &+ C_FC_A  \bigg(
  - \frac{2774476486497511252284870305976724044391}{
    9695938227392717678410570137600000}\noeqnumlb
  & + \frac{1870074094390219597606687}{7855857679623782400}\zeta_3
  - \frac{151838577682432}{136736893668375}L
  - \frac{65536}{188955}L^2
  \bigg)\noeqnumlb
  &+ C_F^2  \bigg(
  - \frac{18248707238521284112217374676684189}{1201215870545458762874880000}\noeqnumlb
  &+ \frac{143085894538571009718455041}{11321677244163686400}\zeta_3
  + \frac{372897204109312}{167122870039125}L
  + \frac{1048576}{230945}L^2
  \bigg)
  \nonumber \end{align}
\begin{align}
  C_{10}^{(2),p} =& + C_FT_R  \bigg(
  \frac{18023374701238546946048}{33879736653956827759125}n_l
  - \frac{1835086571234207640548161}{941510679624652161024000}n_h\noeqnumlb
  &+ \frac{46149597022417}{30236569763840}n_h\zeta_3
  + \frac{2917754077462528}{11642893279392375}L(n_l+n_h)
  + \frac{196608}{1616615}L^2(n_l+n_h)
  \bigg)\noeqnumlb
  &+ C_FC_A  \bigg(
  - \frac{1539378756641920703135316484755839467328419}{1333422361938579459773
    796502732800000}\noeqnumlb
  &+ \frac{12803248135021178751056537}{13331152426028236800}\zeta_3
  - \frac{1179975928287232}{1058444843581125}L
  - \frac{49152}{146965}L^2
  \bigg)\noeqnumlb
  &+ C_F^2  \bigg(
  - \frac{27324314428644557274188086091814971713169}{
    369369075329246387748974100480000}\noeqnumlb
  &+ \frac{13536779710726154580166718023}{219964015029465907200}\zeta_3
  + \frac{423117757374464}{143739423202375}L
  + \frac{7962624}{1616615}L^2
  \bigg)
  \nonumber \end{align}
\begin{align}
  C_{11}^{(2),p} =&  C_FT_R  \bigg(
  \frac{3760136484063862872174592}{7013105487369063346138875}n_l
  - \frac{14436891835130136916846548977}{6626352163198301909286912000}n_h\noeqnumlb
  &+ \frac{186926938565003}{108851651149824}n_h\zeta_3
  + \frac{17577073730093056}{68859397395263475}L(n_l+n_h)
  + \frac{2621440}{22309287}L^2(n_l+n_h)
  \bigg)\noeqnumlb
  &+ C_FC_A  \bigg(
  - \frac{2196670741691390865449882336983942060570342639}{4722982005986448446
    51878721267957760000}\noeqnumlb
  &+ \frac{5140924281484864972357892859}{1328671525127480934400}\zeta_3
  - \frac{6968767912075264}{6259945217751225}L
  - \frac{655360}{2028117}L^2
  \bigg)\noeqnumlb
  &+ C_F^2  \bigg(
  - \frac{19892582218578011106338578696247144162810751553}{562259762617434338
    87128419198566400000}\noeqnumlb
  &+ \frac{821232360279025582272054039829}{2790210202767709962240}\zeta_3
  + \frac{796428906692608}{218601261572265}L
  + \frac{39321600}{7436429}L^2
  \bigg)
  \nonumber \end{align}
\begin{align}
  C_{12}^{(2),p} =&  C_FT_R  \bigg(
  \frac{2361650790215100407548239872}{4384466039694277148308822125}n_l
  - \frac{43084675796375645372731027301693}{17917656249288208362711810048000}
  n_h\noeqnumlb
  &+ \frac{3922441614750337}{2058285767196672}n_h\zeta_3
  + \frac{676219888715104256}{2620413068150662785}L(n_l+n_h)
  + \frac{5767168}{50702925}L^2(n_l+n_h)
  \bigg)\noeqnumlb
  &+ C_FC_A  \bigg(
  - \frac{70034794682581481813249178402809862391078239422033231}{374168992379
    5446447364957401016026833879040000}\noeqnumlb
  &+ \frac{77778913328384051652725192513}{4995061621737998254080}\zeta_3
  - \frac{1319456473035685888}{1191096849159392175}L
  - \frac{15859712}{50702925}L^2
  \bigg)\noeqnumlb
  &+ C_F^2  \bigg(
  - \frac{2267473277094876433590459813785417415724639080638329}{1360218817724
    097152597410717251718348800000}\noeqnumlb
  &+ \frac{66863488712130988880884620697226071}{48214832303826028147507200}\zeta_3
  + \frac{572835921739218944}{132344094351043575}L
  + \frac{95158272}{16900975}L^2
  \bigg)
  \nonumber \end{align}

\begin{align}
  C_{S,9}^{\left(2\right),p}=
  + \frac{27352144660490483390633}{174887254737970790400000}
  + \frac{2454162037509}{9448928051200}\zeta_3
  - \frac{263495168}{2010133125}L_{-q^2}
  \nonumber \end{align}
\begin{align}
  C_{S,10}^{\left(2\right),p}=
  + \frac{99214339418590131445021}{783494901226109140992000}
  + \frac{46177240370863}{199561360441344}\zeta_3
  - \frac{732934144}{6416344935}L_{-q^2}
  \nonumber \end{align}
\begin{align}
  C_{S,11}^{\left(2\right),p}= 
  + \frac{2787996822340736999889074473}{26799913193379798833115955200}
  + \frac{53552348130925}{257285720899584}\zeta_3
  - \frac{21320382464}{211739382855}L_{-q^2}
  \nonumber \end{align}
\begin{align}
  C_{S,12}^{\left(2\right),p}=
  + \frac{14641282147919288479322671871}{169262609642398729472311296000}
  + \frac{561179696637195}{2973079441506304}\zeta_3
  - \frac{1212441755648}{13527793904625}L_{-q^2}
  \nonumber \end{align}

\subsection{Vector Current}
\label{sec:app:vc}
\begin{align}
  \sum_{n=9}^{12} C_n^{\left(0\right),v} z^n = 
  \frac{524288 }{8729721} z^9
  +\frac{524288 }{10140585} z^{10}
  +\frac{8388608 }{185910725} z^{11}
  +\frac{4194304 }{105306075} z^{12}
\nonumber
\end{align}

\begin{align}
  \sum_{n=9}^{12} C_n^{\left(1\right),a} z^n =
&\left(-\frac{2103270735183872}{89816605298
   16975}-\frac{262144
   }{323323}L\right) z^9 \noeqnumlb
&+\left(-\frac{57014372204019712}{2190980826
   21292875}-\frac{524288
   }{676039}L\right) z^{10}\noeqnumlb
&+\left(-\frac{13456105329705877504}{48040
   906249428817725}-\frac{12582912
   }{16900975}L\right) z^{11}\noeqnumlb
& +\left(-\frac{249307596877594624}{8438945
   28756654375}-\frac{8388608
   }{11700675}L\right) z^{12}
\nonumber
\end{align}

\begin{align}
  %\label{tlsc:threeloopvc009}
&C_{9}^{\left(2\right),v}=\bigg(\frac{1769472}{323323} L^2+\frac{1422485887860736}{332654093696925} L\noeqnumlb
&-\frac{381427200524285859827479342961645120691697}{26317546617208805127114404659200000}\noeqnumlb
&+\frac{4641207108486463789099278533}{384937026301565337600} \zeta_3 \bigg) C_F^2+C_A \bigg(-\frac{32768}{88179} L^2-\frac{3201458214330368}{2449543780859175} L\noeqnumlb
&-\frac{4800616671655745922224061798696958246114657}{21001402200532626491437294918041600000}+\frac{62742999460179248975566471}{329946022544198860800} \zeta_3 \bigg) C_F\noeqnumlb
&+T_R \bigg(\bigg(\frac{131072}{969969} n_h+\frac{131072}{969969} n_l\bigg) L^2+\bigg(\frac{8171733794127872}{26944981589450925} n_h\noeqnumlb
&+\frac{8171733794127872}{26944981589450925} n_l\bigg) L-\frac{7505222973653044996167293}{2893884388042736664576000} n_h\noeqnumlb
&+\frac{1716980884029425246777344}{2744258668970503048489125} n_l+\frac{15641077379375}{7653631721472} n_h \zeta_3 \bigg) C_F
\nonumber
\end{align}

\begin{align}
  %\label{tlsc:threeloopvc010}
&C_{10}^{\left(2\right),v}=\bigg(\frac{3932160}{676039} L^2+\frac{72590094055309312}{14606538841419525} L\noeqnumlb
&-\frac{541634465543720570056096925789635688755714501}{7906777911807670390377433949798400000}\noeqnumlb
&+\frac{3503779464918218459698976337}{61482997256500019200} \zeta_3 \bigg) C_F^2\noeqnumlb
&+C_A \bigg(-\frac{720896}{2028117} L^2-\frac{76685205430140928}{59754022533079875} L\noeqnumlb
&-\frac{100571880864387246559861014880105540717546397}{110694890765307385465284075297177600000}\noeqnumlb
&+\frac{358487787916859833012705187}{474297407407285862400} \zeta_3 \bigg) C_F\noeqnumlb
&+T_R \bigg(\bigg(\frac{262144}{2028117} n_h+\frac{262144}{2028117} n_l\bigg) L^2+\bigg(\frac{198611843579379712}{657294247863878625} n_h\noeqnumlb
&+\frac{198611843579379712}{657294247863878625} n_l\bigg) L-\frac{56043284857896604271843338613}{19879056489594905727860736000} n_h\noeqnumlb
&+\frac{1187963065976173458415616}{1912665132918835458037875} n_l+\frac{331575524130137}{148434069749760} n_h \zeta_3 \bigg) C_F
\nonumber
\end{align}

\begin{align}
  %\label{tlsc:threeloopvc011}
&C_{11}^{\left(2\right),v}=\bigg(\frac{103809024}{16900975} L^2+\frac{1643667546316865536}{291157007572295865} L\noeqnumlb
&-\frac{1070887924864138198352946851863557684684339508412651}{3351967800820096554615047838941734502400000}\noeqnumlb
&+\frac{9610834140751883729239482548466517}{36161124227869521110630400} \zeta_3 \bigg) C_F^2\noeqnumlb
&+C_A \bigg(-\frac{5767168}{16900975} L^2
-\frac{16505631011785080832}{13102065340753313925} L\noeqnumlb
&-\frac{10142587654701589960628097458296025022529793100251069}{2806267442846584835523718050762020125409280000}\noeqnumlb
&+\frac{1150547909207069411531224161671}{382657399236714509107200} \zeta_3 \bigg) C_F+T_R \bigg(\bigg(\frac{2097152}{16900975} n_h+\frac{2097152}{16900975} n_l\bigg) L^2\noeqnumlb
&+\bigg(\frac{43261806664333656064}{144122718748286453175} n_h+\frac{43261806664333656064}{144122718748286453175} n_l\bigg) L\noeqnumlb
&-\frac{85142270271685420818460662817}{27996337889512825566737203200} n_h+\frac{849053414353143539189776384}{1377975041046772818039915525} n_l\noeqnumlb
&+\frac{428230928483939}{176883933118464} n_h \zeta_3 \bigg) C_F
\nonumber
\end{align}

\begin{align}
  %\label{tlsc:threeloopvc012}
&C_{12}^{\left(2\right),v}=\bigg(\frac{8388608}{1300075} L^2+\frac{591048256550862848}{93766058750739375} L\noeqnumlb
&-\frac{2120116328634260041790541524636709954701152751347827221}{1439111509152094787448060538852318013030400000}\noeqnumlb
&+\frac{16883124936978831575637459733931549}{13775666372521722327859200} \zeta_3 \bigg) C_F^2+C_A \bigg(-\frac{11534336}{35102025} L^2\noeqnumlb
&-\frac{284624595694944256}{230153053297269375} L-\frac{20238608949682386388157368187133321175530208319894383}{1406650347291521220813893759780461215744000000}\noeqnumlb
&+\frac{103631644405192190650944065895}{8658106811012530307072} \zeta_3 \bigg) C_F+T_R \bigg(\bigg(\frac{4194304}{35102025} n_h+\frac{4194304}{35102025} n_l\bigg) L^2\noeqnumlb
&+\bigg(\frac{753487939399450624}{2531683586269963125} n_h+\frac{753487939399450624}{2531683586269963125} n_l\bigg) L\noeqnumlb
&-\frac{213296190145039197864108250117409}{65438396736530847933382262784000} n_h+\frac{369724970712395017121284096}{605143454135205716089734375} n_l\noeqnumlb
&+\frac{4949453959637945}{1899956092796928} n_h \zeta_3 \bigg) C_F
\nonumber
\end{align}

\subsection{Axial-Vector Current}
%\label{sec:app:av}
% axial vector 1loop and 2loop

\begin{align}
  \sum_{n=9}^{12} C_n^{\left(0\right),a} z^n = \frac{ 262144}{43648605} z^9 +
  \frac{524288}{111546435} z^{10} + \frac{ 2097152}{557732175} z^{11}
  +\frac{4194304}{1368978975} z^{12}
\nonumber
\end{align}

\begin{align}
  \sum_{n=9}^{12} C_n^{\left(1\right),a} z^n =&        \left(
    - \frac{14273234676170752}{314358118543594125}
    - \frac{131072}{1616615}L
  \right)z^9\noeqnumlb
  &+ \left(
    - \frac{101689470330109952}{2410078908834221625}
    - \frac{524288}{7436429}L
  \right)z^{10}\noeqnumlb
  &+ \left(
    - \frac{14155800052957184}{361209821424276825}
    - \frac{1048576}{16900975}L
  \right)z^{11}\noeqnumlb
  & +  \left(
    - \frac{83544812412930228224}{2292861434631829936875}
    - \frac{8388608}{152108775}L
  \right)z^{12}
  \nonumber \end{align}

\begin{align}
  C_{9}^{(2),a} =&C_FT_R  \bigg(
  \frac{209029528206436443461632}{2744258668970503048489125}n_l
  - \frac{721707504396631265775619523}{1296460205843146025730048000}n_h\noeqnumlb&
  + \frac{1102011173546563}{2449162150871040}n_h\zeta_3
  + \frac{35512855382474752}{943074355630782375}L(n_l+n_h)
  + \frac{65536}{4849845}L^2(n_l+n_h)
  \bigg)\noeqnumlb&
  +C_FC_A  \bigg(
  \frac{2044821895603527816981536356704130687706357}{3360224352085220238629
    967186886656000}\noeqnumlb&
  - \frac{668136450989003773285863131}{1319784090176795443200}\zeta_3
  - \frac{12933050525913088}{85734032330071125}L
  - \frac{16384}{440895}L^2
  \bigg)\noeqnumlb&
  +C_F^2  \bigg(
  \frac{217311353924672252345666774738607455908207}{
    5540536129938695816234611507200000}\noeqnumlb&
  - \frac{4567332086889583465526764451}{139977100473296486400}\zeta_3
  + \frac{8434594159026176}{11642893279392375}L
  + \frac{884736}{1616615}L^2
  \bigg)
  \nonumber \end{align}
\begin{align}
  C_{10}^{(2),a} =& C_FT_R  \bigg(
  \frac{287854153300932624846848}{4207863292421438007683325}n_l
  - \frac{209290713314001851931555887}{382289547876825110151168000}n_h\noeqnumlb&
  + \frac{181196559131413}{408193691811840}n_h\zeta_3
  + \frac{243286941705469952}{7230236726502664875}L(n_l+n_h)
  + \frac{262144}{22309287}L^2(n_l+n_h)
  \bigg)\noeqnumlb&
  +C_FC_A  \bigg(
  \frac{8649742991474532665515591016625431163403791601}{3542236504489836334
    889090409509683200000}\noeqnumlb&
  - \frac{44535077782003477491448862123}{21923080164603435417600}\zeta_3
  - \frac{87853979961663488}{657294247863878625}L
  - \frac{65536}{2028117}L^2
  \bigg)\noeqnumlb&
  +C_F^2  \bigg(
  \frac{3633687108507123867092545337354886401661662813}{1946283793675734249
    9390606645657600000}\noeqnumlb&
  - \frac{35752565032549034445184262371577}{230192341728336071884800}\zeta_3
  + \frac{117265192181399552}{160671927255614775}L
  + \frac{3932160}{7436429}L^2
  \bigg)
  \nonumber \end{align}
\begin{align}
  C_{11}^{(2),a} =&C_FT_R  \bigg(
  \frac{157211774390308655582420992}{2538375075612476243757739125}n_l
  - \frac{175561738935918272688195117821}{325775568168876515685669273600}n_h\noeqnumlb&
  + \frac{9918778191302813}{22641143439163392}n_h\zeta_3
  + \frac{32831051515756544}{1083629464272830475}L(n_l+n_h)
  + \frac{524288}{50702925}L^2(n_l+n_h)
  \bigg)\noeqnumlb&
  +C_FC_A  \bigg(
  \frac{1157200807257694296198427317159640234150785132824303}{1181586291724
    87782548367075821558742122496000}\noeqnumlb&
  - \frac{261883214808207508025819129858143}{32143221535884018765004800}\zeta_3
  - \frac{619633587273728}{5184829972597275}L
  - \frac{1441792}{50702925}L^2
  \bigg)\noeqnumlb&
  +C_F^2  \bigg(
  \frac{325018733172338765882468493323383479610810052389621}{37096876847020
    8314344748377432286822400000}\noeqnumlb&
  - \frac{105426049549787146186606803725022377}{144644496911478084442521600}
  \zeta_3
  + \frac{1602332519923712}{2189150432874405}L
  + \frac{8650752}{16900975}L^2
  \bigg)
  \nonumber \end{align}
\begin{align}
  C_{12}^{(2),a} =& 
  C_FT_R  \bigg(
  \frac{92827398267438355820629508096}{1644174764885353930615808296875}n_l
  - \frac{104450315270262159531899064503}{196742892149046993786639482880}n_h\noeqnumlb&
  + \frac{5341389426809173}{12349714603180032}n_h\zeta_3
  + \frac{188918503999998132224}{6878584303895489810625}L(n_l+n_h)\noeqnumlb&
  + \frac{4194304}{456326325}L^2(n_l+n_h)
  \bigg)\noeqnumlb&
  +C_FC_A  \bigg(
  \frac{14691137807907519650408236687148333789997689179014563981}{
    374168992379544644736495740101602683387904000000}\noeqnumlb&
  - \frac{818111578538634207243886625532379}{25046666131857676959744000}\zeta_3
  - \frac{67346421666621587456}{625325845808680891875}L
  - \frac{11534336}{456326325}L^2
  \bigg)\noeqnumlb&
  +C_F^2  \bigg(
  \frac{151997936918859535196394940736545376607752983761243261719}{
    37416899237954464473649574010160268338790400000}\noeqnumlb&
  - \frac{3258783646497863020834421263009179937}{964296646076520562950144000}
  \zeta_3\noeqnumlb&
  + \frac{186408134950156238848}{254762381625758881875}L
  + \frac{8388608}{16900975}L^2
  \bigg)
  \nonumber \end{align}

\begin{align}
  C_{S,9}^{\left(2\right),a}=&   
  - \frac{54729591036013756991855587205067403}{293352919626487481150235230601
    2160000}
  + \frac{24604386555649}{997806802206720}\zeta_3\noeqnumlb
  & - \frac{6722140930048}{997962281090775}L_{-q^2}
  \nonumber \end{align}
\begin{align}
  C_{S,10}^{\left(2\right),a}=&   
  - \frac{119621269712255836267447988400312983}{73604914379009586179513566950
    85056000}
  + \frac{87798501391061}{4245214394843136}\zeta_3\noeqnumlb
  &- \frac{303225290752}{57733355104425}L_{-q^2}
  \nonumber \end{align}
\begin{align}
  C_{S,11}^{\left(2\right),a}=&   
  - \frac{392233656487033430717559964717352391313}{27547265098473410838449785
    029714247680000}
  + \frac{2353323008444009}{133788574867783680}\zeta_3\noeqnumlb
  &- \frac{1219596674400256}{291157007572295865}L_{-q^2}
  \nonumber \end{align}
\begin{align}
  C_{S,12}^{\left(2\right),a}= &  
  - \frac{5473635073785826339847768690355966987576523}{4360943967127559961963
    81981470399397888000000}
  + \frac{26173581704692859}{1728960044445204480}\zeta_3\noeqnumlb
  &- \frac{466514873679872}{137042701251080625}L_{-q^2}
  \nonumber \end{align}

}

%\bibliographystyle{unsrt}
%\bibliography{biblio}

\end{document}